\documentclass[reprint,prc,aps,nofootinbib]{revtex4-2}
\usepackage{geometry}			
\usepackage{graphicx}
\usepackage{float}
\usepackage[T1]{fontenc}
\usepackage[utf8]{inputenc}
\usepackage{amssymb}
\usepackage{amsmath}
\usepackage{graphicx}
\usepackage{comment}
\usepackage{dsfont}
\usepackage{mathrsfs}
\usepackage{anyfontsize}
\usepackage{natbib}
\usepackage{url}
\usepackage{slashed}
\newcommand{\chiPT}{$\chi${PT }}

\newcommand{\SUtw}{{SU(2)}}
\newcommand{\SUtwLR}{{SU(2)$_L\times$SU(2)$_R$ }}
\newcommand{\chiPTtw}{\SUtw$\chi$PT}
\newcommand{\chiPTtwb}{\SUtw$\chi$PT$\;$}

\newcommand{\chiNL}{$\chi$NL}
\newcommand{\SchiNL}{Static\chiNL} 
\newcommand{\SchiNLb}{Static\chiNL\;}
\newcommand{\stimes}{\,\!{\times}\!\,}
\newcommand{\half}{\frac{1}{2}}
\newcommand{\shalf}{\tfrac{1}{2}}
\newcommand{\bra}{\big< \chi NL \big\vert}
\newcommand{\ket}{\big\vert \chi NL \big>}

\newcommand{\bran}{\big<}  
\newcommand{\ketn}{\big>}  

\newcommand{\tildemone}{m^N\mathds{1}}
\newcommand{\tildemn}{{m}_n}

\newcommand{\calC}{{\mathcal C}}

\DeclareMathOperator{\Tr}{Tr}
\DeclareMathOperator{\expon}{exp}

\newcommand{\Lpionless}{\Lambda_{\pi\mkern-9.5mu/}}

\newcommand{\be}{\begin{equation}}
\newcommand{\ee}{\end{equation}}
\newcommand{\ba}{\begin{aligned}}
\newcommand{\ea}{\end{aligned}}
\newenvironment{nalign}{
    \begin{equation}
    \begin{aligned}
}{
    \end{aligned}
    \end{equation}
    \ignorespacesafterend
}
\newcommand{\bna}{\begin{nalign}}
\newcommand{\ena}{\end{nalign}}

\begin{document}
\title{Nuclear matter as a liquid phase of spontaneously broken
semi-classical \SUtwLR chiral  perturbation theory:\\
Static chiral nucleon liquids
}
\author{Bryan W. Lynn}
\thanks{Deceased}
\author{Brian J. Coffey}
\affiliation{Dept Physics/CERCA/ISO, CWRU, Cleveland, OH, 44106 USA}
\thanks{E-mail: bjc118@case.edu}
\author{Kellen E. McGee}
\affiliation{Dept Physics/Astronomy, Michigan State U., East Lansing MI, 48824}
\email{E-mail: mcgee@frib.msu.edu}
\author{Glenn D. Starkman}
\affiliation{Dept Physics/CERCA/ISO, CWRU, Cleveland, OH, 44106 USA}
\affiliation{Dept Astronomy, CWRU, Cleveland, OH, 44106 USA}
\email{E-mail: glenn.starkman@case.edu}
\date{\today}
\setlength{\oddsidemargin}{11pt}
\setlength{\textwidth}{480pt}
\newgeometry{vmargin={25mm}, hmargin={25mm,25mm}}

\begin{abstract}
The Standard Model of particle physics (SM), augmented with neutrino mixing, is either
the complete theory of interactions of known particles at energies accessible to Nature on Earth,
or very nearly so.  Starting with a Lagrangian symmetric under the global \SUtwLR symmetry of two-massless-quark QCD, {\it spontaneously} broken to SU$(2)_{L+R}$.  Using naive dimensional operator power counting that enables perturbation and truncation
in inverse powers of $\Lambda_{\chi SB}\sim 1 GeV$, we show that, to ${\cal O}(\Lambda_{\chi SB})$ and ${\cal O}(\Lambda^0_{\chi SB})$, \SUtw $\chi$PT of protons, neutrons and pions admits a liquid 
phase, with energy required to increase or decrease the nucleon density.
We further show that  in the semi-classical approximation -- i.e., quantum nucleons and classical pions -- 
"Pionless \chiPTtw'' emerges in that chiral liquid: soft static infrared Nambu-Goldstone-Boson pions decouple from 
``Static Chiral Nucleon Liquids" (\SchiNL).
This vastly simplifies the derivation
of saturated nuclear matter (the infinite liquid phase) 
and of finite microscopic liquid drops (ground-state heavy nuclides).
\SchiNLb
are 
made entirely of nucleons.  
They have even parity, total spin zero,  
even proton number $Z$ and  even neutron number $N$.
The nucleons are arranged so 
local expectation values for spin and momentum
vanish.
We derive the 
\SchiNLb effective Lagrangian from semi-classical \chiPTtwb symmetries 
to order $\Lambda_{\chi SB},\Lambda^0_{\chi SB}$ including:
all  relativistic 4-nucleon operators 
that survive Fierz rearrangement in the non-relativistic limit;
\chiPTtwb fermion exchange operators and 
iso-vector exchange 
operators which are 
important when $Z\neq N$. 

Mean-field  \SchiNLb non-topological solitons are true solutions of \chiPTtwb semi-classical symmetries: e.g. they obey all CVC, PCAC conservation laws. They have zero internal and external pressure. The nuclear liquid-drop model and 
Bethe-von Weizs\"acker semi-empirical mass formula emerge -- with correct nuclear density and saturation and asymmetry energies -- in an
explicit Thomas-Fermi construction. 
\end{abstract} 

\maketitle

\section{Introduction}
\label{Introduction}
In the Standard Model (SM) of particle physics, 
Quantum Chromodynamics (QCD) 
describes the strong interactions among quarks and gluons. 
At low energies, quarks and gluons are confined inside hadrons, 
concealing their degrees of freedom in such a way that 
we must employ an effective field theory (EFT) of hadrons. 
In doing so, we acknowledge as a starting point 
a still-mysterious experimental fact: 
Nature first makes hadrons 
and then assembles nuclei from them \cite{Ericson1988, Weinberg1990288, Ordonez1992.459, vanKolck2002191}.
\par 
Since nuclei are made of hadrons, 
the fundamental challenge of nuclear physics is to
identify the correct EFT of hadrons
and use it to characterize all nuclear physics observations.  (See the recent review by Hammer et al. \cite{Hammer2020.025004}.)
Ultimately, the correct choice of EFT will both match the observations
and be derivable from the SM, i.e., QCD. 

Chiral perturbation theory (\chiPT) \cite{Weinberg19681568, Weinberg1978.327, Coleman19692239, Callan19692247, Gasser1984142, Gasser1985465} 
is a low-energy perturbative approach to identifying 
the operators in the EFT of hadrons 
that are allowed by the global symmetries of the SM.
It builds on the observation that the up and down quarks 
($m_{u}\simeq 6$ MeV, $m_{d}\simeq 12$ MeV),
as well as the 3 pions ($\pi^\pm,\pi^0$, $m_{\pi}\simeq 140$ MeV)--which are pseudo-Nambu-Goldstone bosons (pNGBs) of the chiral symmetry--are all nearly massless compared to the cut-off energy scale in low-energy hadronic physics
$\Lambda_{\chi SB} \sim  1GeV$.

With naive power counting \cite{Manohar1984189}, 
the effective Lagrangian  of \SUtwLR 
\chiPT~
incorporates explicit breaking.  
The resultant perturbation expansion 
in the inverse of the chiral-symmetry-breaking scale 
$\Lambda^{-1}_{\chi SB}\sim 1$ GeV$^{-1}$ 
renders \chiPTtw's strong interaction predictions calculable in practice. 
Its low-energy dynamics of a proton-neutron nucleon doublet 
	and three pions as a pNGB triplet
are our best understanding, together with lattice QCD, of the experimentally observed low-energy dynamics of QCD strong interactions.  The predictive power of \chiPT  
\cite{Weinberg19681568, Coleman19692239, Callan19692247, Gasser1984142, Gasser1985465, Manohar1984189, Georgi1984, Borasoy199885, Jenkins1991113}
derives from its ability to maintain a well-ordered low-energy perturbation expansion 
that can be truncated. 

B.W. Lynn \cite{Lynn1993281} first introduced the idea that \chiPTtwb could also admit a liquid phase and introduced the idea of an `\SUtwLR chiral liquid' as a statistically significant number of baryons interacting via chiral operators with an almost constant saturated density which can can survive as localized liquid drops at zero external pressure.  The Lagrangian included all analytic \chiPTtwb terms of ${\mathcal O}(\Lambda_{\chi SB})$ and ${\mathcal O}(\Lambda_{\chi SB}^0)$.
Lynn argued that, in the exact chiral limit, nucleons in the liquid phase interact with each-other only via the contact terms in \eqref{BosonExchangeContactInteractions}. 
Study of chiral liquids in \cite{Lynn1993281} focused on those explicit chiral symmetry breaking terms whose origin lies entirely in the non-zero light quark masses.

The result is a semi-classical nuclear picture, where Thomas-Fermi nucleons with contact interactions move in a mean spherically symmetric ``classical'' pion field, 
which in turn generates a ``no-core'' radial potential for nucleons.
Finite saturating heavy nuclei, with well-defined surfaces, emerge as microscopic droplets of chiral liquid.
Saturating infinite nuclear matter emerges as very large drops of chiral liquid, while neutron stars (Q-Stars) emerge as oceans of chiral liquid:
These droplets emerge as non-topological-soliton semi-classical solutions of explicitly-broken \chiPTtw.  Lynn \cite{Lynn1993281} conjectured the possible emergence of shell structure in that no-core spherical potential based on the observation that the angular momentum of each nucleon is a good quantum number.
Ref. \cite{Lynn1993281} did not derive semi-classical pionless \chiPTtw.  Here, we focus our study of chiral liquids in the chiral limit, and prove the emergence of semi-classical pionless \chiPTtwb solutions.

There is a long history of viewing nuclear matter as a non-topological soliton.  In the mid 1970's T.D.~Lee and co-workers
\cite{Lee19742291,
Lee1975267, Lee19751591},
S.A. Chin \& J.D. Walecka \cite{Chin197424}, 
and R. Serber \cite{Serber1976718}
first identified 
certain fermion non-topological solitons 
with the ground state of heavy nuclei (as well as possible super-heavy nuclei) in "normal" and "abnormal" phases,
thus making a crucial connection to the ancient (but still persistently predictive) insight of nuclear liquids, 
such as G. Gamow's nuclear liquid-drop model (NLDM) 
and H. Bethe \& C.F. von Weizs\"acker's semi-empirical Mass-Formula (SEMF).  Breaking all precedent, these workers proposed for the first time a theory of liquid nuclear structure composed entirely of nucleons and a static scalar field, with no pions! 

Mathematically,  
such solutions emerge 
as a sub-species of non-topological solitons or Q-balls \cite{Bahcall199067, Bahcall1989606, Selipsky1989430, Lynn1989465, Bahcall1998959, Rosen1968999, Lee19742291, Lee1976254, Friedberg19762739,  Friedberg197632, Coleman1985263, Friedberg19771694, Werle1977367, Morris1978337, Morris197887}, 
a certain sub-set of which are composed of fermions along with the usual scalars.  
A practical goal was to identify mean-field nucleon non-topological solitons 
with the ground state of ordinary even-even spin-zero spherically symmetric heavy nuclei, 
such as $^{40}_{20}$Ca, $_{40}^{90}$Zr, 
	and $^{208}_{\,\,82}$Pb. 

Nuclear non-topological solitons identified as nuclear liquids became popular  with the 
work of Chin \& Walecka \cite{Chin197424}, carried forward by Serot \cite{Serot1979172}.
Walecka's nuclear Quantum Hadrodynamics -1 (QHD-1)  models \cite{Chin1977301, Serot19861, Serr197810} 
contain four dynamical particles: 
protons, neutrons, the Lorentz-scalar iso-scalar $\sigma$, 
and the Lorentz-vector iso-scalar $\omega_\mu$.  Nucleons are treated as 
	locally free-particles in Thomas-Fermi approximation. 
Finite-width nuclear surfaces are generated 
	by dynamical attractive $\sigma$-particle exchange, 
allowing them to exist at zero external pressure.

The empirical success of QHD-1 is based on balancing $\sigma$-boson-exchange attraction
against $\omega_\mu$-boson-exchange repulsion. 
That that balance must be fine-tuned 
remains a famous mystery of the structure of the QHD-1 ground state.
In the absence of long-ranged electromagnetic forces, 
infinite symmetric  $Z=N$ nuclear matter, 
as well as finite microscopic ground state $Z=N$ nuclides, 
appear as symmetric nuclear liquid drops.
\noindent
These nuclear non-topological solitons of are to be classified as liquids because:
\begin{itemize}
\item they have no crystalline 
		or other solid structure;
\item it costs energy to either increase 
		or decrease the density 
		of the constituent nucleons 
		compared to an optimum value;
\item they survive at zero external pressure, 
		e.g. in the absence of gravity, 
		so they are not a ``gas.''
\end{itemize}


Despite their successes, 
such topological soliton models suffer from the flaw that higher loop corrections do not necessarily decrease in size and importance which can significantly renormalize the parameters at each order.  This was first demonstrated by Furnstahl et al. \cite{Furnstahl1989.321} for the Walecka model in the two-loop case.  (See also the discussion in \cite{Furnstahl1997.446}.)

This paper cures those problems, and resurrects nuclear liquids as a good starting point toward understanding the properties of bound nuclear matter 
(with $Z$ and $N$ both even) 
by strict compliance with the requirements of \chiPTtwb effective field theory
of protons, neutrons and pions. 
The static chiral nucleon liquids (\SchiNL) studied below are true solutions to semi-classical \chiPTtwb, 
and have all of the semi-classical symmetries of spontaneously broken \chiPTtwb found in Appendix \ref{app:chiptLagrangian}: they obey all CVC and PCAC Ward identities; 
they are dependent on just a few 
experimentally measurable chiral coefficients; 
and, 
by the symmetries of spontaneously broken \chiPTtw, 
they restore (cf. Appendix A) theoretical predictive power over heavy nuclides.

\section{The emergence of semi-classical pionless \SchiNLb}
In this paper we focus on the chiral limit and postpone treatment of departures from the chiral limit to future work.
The \chiPTtwb Lagrangian
with all terms of order 
$\Lambda_{\chi SB}$ and $\Lambda^0_{\chi SB}$ 
in the chiral limit is:
\begin{nalign}
\label{Symmetric_Lagrangian}
L^{Sym}_{\chi PT} =&L^{\pi; Sym}_{\chi PT} + L^{N; Sym}_{\chi PT} + L^{4-N;Sym}_{\chi PT} \\
L^{\pi; Sym}_{\chi PT} 
=& \frac{f^2_\pi}{4}\Tr \partial_\mu \Sigma \partial^\mu \Sigma^\dagger  \\
L^{N; Sym}_{\chi PT} 
=& \overline{N} \left(i\gamma^\mu (\partial_\mu + V_\mu) -  \tildemone \right)N  \\
&- g_A\overline{N}\gamma^\mu\gamma^5{A_\mu N }  \\
=& \overline{N} \left(i\gamma^\mu \partial_\mu -  \tildemone \right)N 
+i{\vec J}^\mu \cdot{\vec V}_\mu \\
&-g_A {\vec J}^{\mu,5}\cdot {\vec A}_\mu \\
L^{4-N;Sym}_{\chi PT} 
=& C_{\mathscr A} \frac{1}{2f^2_\pi}( \overline{N}\gamma^{\mathscr A} N)( \overline{N}\gamma_{\mathscr A} N) 
+++,
\end{nalign}
where the pion field is:
\be
\label{Sigmarepn_main}
\Sigma \equiv \expon(2i \pi_a \frac{t_a}{f_\pi})\,,
\ee
and we defined the fermion bi-linear and pionic currents:
\be\ba
{\vec J}^\mu &=\overline{N}\gamma^\mu {\vec t}N,  \quad {\vec J}^{\mu, 5} =\overline{N}\gamma^\mu \gamma^5 {\vec t}N, \\
{\vec J}^\mu &=\overline{N}\gamma^\mu {\vec t}N,  \quad {\vec J}^{\mu, 5} =\overline{N}\gamma^\mu \gamma^5 {\vec t}N, \\
V_\mu&={\vec t}\cdot {\vec V}_\mu\,,\,\,\,
{\vec V}_\mu =  2i~\mathrm{sinc}^2\left(\frac{{\pi}}{2f_\pi}\right)
\left[{\vec \pi}\times \partial_\mu {\vec \pi}\right],  \\
A_\mu&={\vec t}\cdot {\vec A}_\mu, \\
		\quad
		{\vec A}_\mu &= \\
&\!\!\!\!\!			-\frac{2}{{ \pi}^2}
			\left[ {\vec \pi}\left({\vec \pi}\cdot \partial_\mu {\vec \pi}\right) 
			+ \mathrm{sinc}\bigl(\frac{{\pi}}{f_\pi}\bigr) 
			  \left({\vec \pi}\times\left(\partial_\mu{\vec \pi}\times {\vec \pi}\right)\right)\right], 
\ea\ee
with $\vec{t} \equiv \tfrac{1}{2} \vec{\tau}$, $\tau$ are the Pauli iso-spin matrices, $\pi = \vert{\vec \pi}\vert =\sqrt{{\vec \pi}^2}$, 
and $\mathrm{sinc}(x)\equiv\sin(x)/x$.
The pion$\rightarrow$di-leptons decay constant is $F_\pi = 130.4\pm 0.04 \pm 0.2$~MeV \cite{Amsler20081}.  
We use $f_\pi \equiv {F_\pi}/{\sqrt{2}} = 92.2$ MeV.

The parentheses in the four-nucleon Lagrangian 
indicate the order of SU$(2)$ index contraction,
while  $+++$  indicates that one should include 
all possible combinations of such contractions.
As usual,
  $\gamma^{\mathscr A} \equiv
  \left(1, \gamma^\mu,  i\sigma^{\mu\nu},  
  i\gamma^\mu \gamma^5,  \gamma^5\right)$,
for ${\mathscr A}=1,...,16$
(with 
$\sigma^{\mu\nu} \equiv \frac{1}{2}[\gamma^\mu,  \gamma^\nu ]$).
These are commonly referred to as 
scalar (S), vector (V), tensor (T), axial-vector (A), and pseudo-scalar (P)
respectively.
$C_{\mathscr A}$ are a set of chiral constants.


In the chiral limit, where $\vec \pi$'s are massless, 
the presence of quantum nucleon sources 
could allow the massless NGB to build up, 
with tree-level interactions only, 
a non-linear quantum pion cloud. 
If we minimize the resultant action with respect to variations in the pion field, the equations of motion\footnote{
	This is a chiral-limit \chiPTtwb analogue of QED where, in the presence of quantum lepton sources, a specific superposition of massless infra-red photons builds up into a classical electromagnetic field. 
	Important examples are the ``exponentiation" of IR photons in $e^+e^-\rightarrow \mu^+\mu^-$ asymmetries, and $e^+e^-\rightarrow e^+e^-$ Bhabha scattering, at LEP1. 
	Understanding the classical fields generated by initial-state and final-state soft-photon radiation 
	\cite{Jadach1990229, Yennie1961.379}   
	is crucial to dis-entangling high-precision electro-weak loop effects, such as the experimentally confirmed precise Standard Model predictions for the 
top-quark \cite{Lynn1984} and Higgs' masses \cite{Lynn1984, Djouadi1987265}.
} 
capture the part of the quantum cloud 
that is to be characterized as a classical soft-pion field, 
thus giving us the pion ground state 
in the presence of the ground state 
``Chiral Nucleon Liquid" (\chiNL) 
with fixed baryon number $A=Z+N$:
\bna
0=&\left[\partial_\nu \frac{\partial}{\partial \left(\partial_\nu\pi^m\right)}-\frac{\partial}{\partial \pi^m} \right] L_{\chi PT}^{\pi ;Sym} \\
&+ i{\vec J}^\mu \cdot \left[\partial_\nu \frac{\partial}{\partial \left(\partial_\nu\pi^m\right)}
-\frac{\partial}{\partial \pi^m} \right]{\vec V}_\mu \\
&-g_A {\vec J}^{\mu,5}\cdot \left[\partial_\nu \frac{\partial}{\partial \left(\partial_\nu\pi^m\right)}-\frac{\partial}{\partial \pi^m} \right]{\vec A}_\mu  \\
&- 2  \partial_\mu {\vec J}^\mu \cdot 
\mathrm{sinc}^2\left(\frac{{\pi}}{2f_\pi}\right)
\left({\vec \pi}\times {\hat m}\right)
\\
&+\frac{2}{{\pi}^2} g_A \partial_\mu{\vec J}^{\mu,5}\cdot \\
&\times \left[ {\vec \pi}\left({\vec \pi}\cdot {\hat m}\right) 
+ \mathrm{sinc}\left(\frac{{\pi}}{f_\pi}\right)
\left({\vec \pi}\times\left({\hat m}
\times {\vec \pi}\right)\right)\right].
\ena
We divide the classical pion field into 
 ``IR" and ``non-IR" parts.
By definition, only IR pions survive the internal projection operators 
associated with taking expectation values 
of the classical NGB $\vec \pi\,$s 
in the $\ket$ quantum state:
\be
\ba
\label{IRPions}
\bra&  F\left(\partial_\mu {\vec \pi}, {\vec \pi} \right)\ket \\
&= \bra \mathrm{IR\!-\!part}\left[F\left( \partial_\mu {\vec \pi}, {\vec \pi} \right)\right]\ket \\
&\equiv \left\{F\left( \partial_\mu {\vec \pi}, {\vec \pi} \right) \right\}_{IR}, \\
\ea\ee
where $F$ is an unspecified function.  The IR part does not change the \chiNL.
It could in principle play an important role in the excited states  
of the \chiNL: 
	a $\vec \pi$ condensate,
	a giant resonance,
	a breathing mode,
	or a time-dependent flashing-pion mode.
To ignore such classical IR $\vec \pi\,$s 
would therefore be an incorrect definition of the excited states of \chiNL.

We call these ``IR pions" by keeping in mind a simple picture, 
where the $\vec \pi$ wavelength is 
longer than the scale within the \chiNL~  
	over which the local mean values of nucleon spin and momentum vanish.
Only such IR pions survive the internal projection operators associated with taking expectation values of the classical NGB $\vec \pi\,$s in the $\ket$ quantum state.

We now take expectation values of the $\vec \pi$ equations of motion. 
In the presence of the quantum \chiNL~ source,
the classical NGB $\vec \pi$ cloud obeys
\begin{widetext}
\be\ba
 \label{EoMExpectation}
 0&=\bra \left[\partial_\nu \frac{\partial}{\partial \left(\partial_\nu\pi^m\right)} -\frac{\partial}{\partial \pi^m} \right] L_{\chi PT}^{Sym} \ket \\
&=\left\{ \left[\partial_\nu \frac{\partial}{\partial \left(\partial_\nu\pi^m\right)} -\frac{\partial}{\partial \pi^m} \right] L_{\chi PT}^{\pi ;Sym}\right\}_{IR} + i \bra {\vec J}^\mu \ket \cdot \left\{\left[\partial_\nu \frac{\partial}{\partial \left(\partial_\nu\pi^m\right)} -\frac{\partial}{\partial \pi^m} \right]{\vec V}_\mu\right\}_{IR} \\
&-g_A \bra {\vec J}^{\mu,5} \ket \cdot \left\{\left[\partial_\nu \frac{\partial}{\partial \left(\partial_\nu\pi^m\right)} -\frac{\partial}{\partial \pi^m} \right]{\vec A}_\mu\right\}_{IR}  \\
&-  2 \bra \partial_\mu {\vec J}^\mu \ket \cdot 
	\left\{
	\mathrm{sinc}^2\left(\frac{{\pi}}{2f_\pi}\right)
	{\vec \pi}\times {\hat m} 
	\right\}_{IR} \\
&+ \frac{2}{{ \pi}^2} g_A \bra \partial_\mu{\vec J}^{\mu,5} \ket \cdot \left\{{\vec \pi}\left({\vec \pi}\cdot {\hat m}\right)
+\mathrm{sinc}\left(\frac{{\pi}}{f_\pi}\right)
{\vec \pi}\times\left({\hat m}\times {\vec \pi}\right)\right\}_{IR}.
\ea\ee
\end{widetext}
\noindent
We examine the following semi-classical nuclear current components:
\be
\ba
\label{Nucleon_bilinears}
J^\mu_\pm &=J^\mu_1 \pm iJ^\mu_2 = \left \{ \begin{tabular}{c} $\overline{p}\gamma^\mu n$ \\ $\overline{n}\gamma^\mu p$ \end{tabular} \right\}; \\
J^\mu_3 &= \frac{1}{2}\left ( \overline{p}\gamma^\mu p - \overline{n}\gamma^\mu n \right);  \\
J^{5\mu}_\pm &= J^{5\mu}_1 \pm iJ^{5\mu}_2 = \left \{ \begin{tabular}{c} $\overline{p}\gamma^\mu \gamma^5 n$ \\ $\overline{n}\gamma^\mu  \gamma^5 p$ \end{tabular} \right\}; \\
J^{5\mu}_{3} &= \frac{1}{2}\left ( \overline{p}\gamma^\mu \gamma^5 p - \overline{n}\gamma^\mu \gamma^5 n \right);
\ea
\ee
and find that the ground-state expectation values of these currents
and their divergences in \eqref{EoMExpectation} vanish:
\bna
\label{VanishingCurrents1}
\bra {J}_{\mu}^\pm \ket &=\bra {J}_{\mu}^{\pm,5} \ket =0,\\
\bra \partial^\mu{J}_{\mu}^\pm \ket &=\bra \partial^\mu{J}_{\mu}^{\pm,5} \ket =0,
\ena
because ${J}_{\mu}^\pm$ and ${J}_{\mu}^{\pm,5}$ change neutron and proton number.
Since the liquid ground state is homogeneous, isotropic and spherically symmetric,
spatial components of vector currents vanish, in particular
\begin{eqnarray}
\label{VanishingCurrents3}
\bra {J}_{i}^3 \ket \simeq 0\
\end{eqnarray}
for Lorentz index $i\!\!=\!\!1,2,3$.
Because 
there are separately equal numbers of left-handed and right-handed protons and neutrons in the nuclear ground state we have:
\begin{eqnarray}
\label{VanishingCurrents4}
\bra {J}_{\mu}^{3,5} \ket \simeq 0\
\end{eqnarray}
for all $\mu$.  Note that \eqref{VanishingCurrents1}-\eqref{VanishingCurrents4} follow because the liquid ground state is assumed to have definite numbers of fully paired nucleons in a spherically symmetric, homogeneous, and isotropic arrangement.
Current conservation enforces 
\be
\ba
\label{VanishingCurrents5}
\bra \partial^\mu{J}_{\mu}^{3} \ket= 
\bra \partial^\mu{J}_{\mu}^{3,5} \ket  = 0,
\ea
\ee
which leaves only a single non-vanishing current expectation value:
\begin{eqnarray}
\label{VanishingCurrents}
\bra {J}_{0}^3 \ket &\neq& 0 \,.
\end{eqnarray}
\noindent
Equation \eqref{EoMExpectation},  
governing the classical pion cloud,
is thus enormously simplified
\be\ba
0 &\simeq \left\{
\left[\partial_\nu \frac{\partial}{\partial \left(\partial_\nu\pi^m\right)} - \frac{\partial}{\partial \pi^m} \right] L_{\chi PT}^{\pi ;Sym}\right\}_{IR} \\
&+\!i \bra {J}^{3;0} \ket \left\{\left[\partial_\nu \frac{\partial}{\partial \left(\partial_\nu\pi^m\right)} \!-\!\frac{\partial}{\partial \pi^m} \right]{V}_0^3 \right\}_{IR}
\ea\ee
with
\begin{widetext}
\be\ba
\label{SoftPionEnergy}
&\left\{\left[\partial_\nu \frac{\partial}{\partial \left(\partial_\nu\pi^m\right)} -\frac{\partial}{\partial \pi^m} \right]{V}_0^3\right\}_{IR}
=\\
&\quad \qquad\left\{
2i\left[\left(\partial_0{\vec \pi}\right)\times {\hat m}+{\vec \pi}\times {\hat m}\partial_0 -{\hat m}\times \left(\partial_0{\vec \pi}\right)  
	-{\vec \pi}\times \left(\partial_0{\vec \pi}\right)\frac{\partial}{\partial \pi^m}  \right]^3 \mathrm{sinc}^2\left(\frac{{\pi}}{2f_\pi}\right) \right\}_{IR}.
\ea\ee
\end{widetext}

A crucial observation is that \eqref{SoftPionEnergy} 
is linear in $\partial_0{\vec \pi}$;
i.e., in the energy of the classical NGB IR $\vec \pi$ field.
Expecting the nuclear ground state, 
and thus its  classical IR $\vec\pi$ field, 
to be static, we enforce 
\begin{equation}
\label{StaticPion}
\left\{\partial_o{\vec \pi}  \right\}_{IR} = 0\,.
\end{equation}
It now follows that 
\begin{eqnarray}
\label{PionDecouplinga}
 \left\{\left[\partial_\nu \frac{\partial}{\partial \left(\partial_\nu\pi^m\right)} -\frac{\partial}{\partial \pi^m} \right]{V}_0^3 \right\}_{IR} &=&0 \,,
\end{eqnarray}
independently of $\bra {J}^{3;0} \ket$. 
The IR pion equation of motion
\begin{eqnarray}
\label{PionDecouplingb}
  \left\{
 \left[\partial_\nu \frac{\partial}{\partial \left(\partial_\nu\pi^m\right)} - \frac{\partial}{\partial \pi^m} \right] L_{\chi PT}^{\pi ;Sym}\right\}_{IR} &=&0 \,,
\end{eqnarray}
therefore has no nucleon source.
The ground state nucleons are not a source of any 
static IR NGB $\vec \pi$ classical field.  The nuclear ground state in the chiral liquid is thus a static chiral nucleon liquid 
(\SchiNL), with no $\vec \pi$ condensate\footnote{
	After explicit chiral symmetry breaking, 
	with non-zero $u,d$ quark and resultant pion masses, 
	and with Partially Conserved  Axial Currents (PCAC), 
	a static  S-wave $\vec \pi$ condensate 
	is a logical possibility \cite{Lynn1993281}. 
}
or time-dependent pion-flashing modes.  


We want to quantize the nucleons in the background field of the static \chiNL, 
and so consider the expectation value of the nucleon equation of motion
in the chiral nucleon liquid static ground state.  For brevity, we denote expectations in this ground state using '$\bran$' and '$\ketn$':
\be
\ba
\label{ChiralLiquidDiracEquation}
0=& \bran \overline{N} \frac{\partial}{\partial \overline{N} }L^{Sym}_{\chi PT} \ketn \\
=& \bran \overline{N} \left(i\gamma^\mu \partial_\mu -  \tildemone \right)N \ketn \\
&+ i \bran{\vec J}^\mu\ketn \cdot \left\{ {\vec V}_\mu \right\}_{IR} -g_A\bran {\vec J}^{\mu,5}\ketn \cdot \left\{ {\vec A}_\mu  \right\}_{IR} \\
&+ \frac{1}{f^2_\pi}\bran C_{\mathscr A}( \overline{N}\gamma^{\mathscr A} N)( \overline{N}\gamma_{\mathscr A} N) +++ \ketn\,.
\ea
\ee
Since
most of the nucleon \SUtwLR currents 
vanish in the \SchiNL,
and since $\left\{\partial_o{\vec \pi}  \right\}_{IR} = 0$, we find:
\begin{eqnarray}
\label{Static_chiral_liquid_dirac_equation}
0 &\simeq& \bran \overline{N} \left(i\gamma^\mu \partial_\mu -  \tildemone \right)N \ketn \\
&&+ \frac{1}{f^2_\pi}\bran C_{\mathscr A}( \overline{N}\gamma^{\mathscr A} N)( \overline{N}\gamma_{\mathscr A} N) +++ \ketn\,.
\end{eqnarray}
\noindent
Equations \eqref{PionDecouplingb} 
	and \eqref{Static_chiral_liquid_dirac_equation} 
show that, 
to order  $\Lambda_{\chi SB}$ and $\Lambda^0_{\chi SB}$,
\SchiNLb are composed entirely of nucleons. 
That is also the basic premise of many empirical models 
and we have shown that that empirical nuclear premise 
can be (to good approximation) traced directly 
to the global \SUtwLR symmetries 
	of 2-massless-quark QCD of the Standard Model.

The effective Lagrangian
derived from \SUtwLR $\chi$PT 
governing \SchiNLb can now be written:
\be
\ba
\label{NucleonLiquidLagrangian}
\bran L^{Sym}_{\chi PT}\ketn &\equiv L_{S\chi NL} \\
L_{S\chi NL} &= L^{Free}_{S\chi NL}  +L^{4-N}_{S\chi NL} \\
L^{Free}_{S\chi NL} &= \bran \overline{N} \left(i\gamma^\mu \partial_\mu -  \tildemone \right)N   \ketn \\
L^{4-N}_{S\chi NL} &= \bran \frac{1}{2f^2_\pi}C_{\mathscr A} ( \overline{N}\gamma^{\mathscr A} N)( \overline{N}\gamma_{\mathscr A} N) +++ \ketn.
\ea
\ee

Semi-classical pionless \chiPTtwb thus emerges inside Static\chiNL.
Within all-loop-orders renormalized analytic \chiPTtwb to ${\cal O}(\Lambda_{\chi SB})$ and ${\cal O}(\Lambda^0_{\chi SB})$,
infrared NGB pions effectively decouple from \SchiNL, 
vastly simplifying the derivation 
of the properties  of saturated nuclear matter 
(the infinite liquid phase) 
and of finite microscopic liquid drops (the nuclides). 
\SchiNLb thus explain 
the (previously puzzling) power of pionless EFT 
to capture experimental ground state facts of certain specific nuclides, 
by tracing that empirical success 
directly to the global symmetries of two-massless-quark QCD.

It will be shown below that static \chiNL s 
satisfy all relevant \SUtwLR
vector and axial-vector current-conservation equations 
in the liquid phase.  
\SchiNLb are therefore solutions 
of the semi-classical-liquid equations of motion, possessing the symmetries of spontaneously broken \chiPTtwb (cf. Appendix A.1).

\section{Semi-classical pionless \SchiNLb as the approximate ground state of certain nuclei}
To further elucidate the properties of the static \chiNL,
we must address the effects of the four-nucleon interactions.
In this paper, we ignore fluctuations in all bi-linear nucleon operators.  For our purposes this is equivalent to ignoring any and all nuclear excited states.

{\it A priori} there are 10 possible contact interactions representing
isosinglet and isotriplet  channels for each of five spatial  current types:
scalar, vector, tensor, pseudo-scalar and axial-vector.
There are therefore ten chiral coefficients parametrizing 4-nucleon contact terms:
$C^{T}_{K}$ with $K\in\{S,V,T,A,P\}$ and $T\in\{0,1\}$.

The inclusion of exchange interactions 
induces the isospin ($T=1$) operators 
	to appear \cite{Lynn1993281}, 
and potentially greatly complicates 
	the effective chiral Lagrangian.
Fortunately, we are interested here 
	in the liquid limit of this Lagrangian.
Spinor-interchange contributions 
	are properly obtained by Fierz rearranging before 
	imposing the properties of the semi-classical liquid (see Appendix B).
The appropriate \SchiNLb Lagrangian, is given by
\begin{eqnarray}
\label{LagrangianMF-StaticXNL}
L_{S\chi NL} &=& {\bar N}\left( i\gamma^\mu \, \overrightarrow\partial_\mu +\Theta\right) N + L_{S\chi NL}^{4-N;BE},
\end{eqnarray}
where the contact interactions can be approximated by:
\be\ba
\label{BosonExchangeContactInteractions}
-&L_{S\chi NL}^{4-N;BE}=\frac{C^{S}_{200}}{2f^2_\pi} \left< {\overline N}N \right> \left< {\overline N}N \right> \\
&- \frac{{\overline {C^{S}_{200}}}}{4f^2_\pi}   \left\{ \left< {\overline N}N \right> \left< {\overline N}N \right> 
+ 4\left< {\overline N}t_3 N \right> \left< {\overline N}t_3 N \right> 
 \right\} \\
&+\frac{\calC^{V}_{200}}{2f^2_\pi}  \left\{  \left< {N^{\dagger}}N \right> \left< {N^{\dagger}}N \right>  \right\}\\
&- \frac{{\overline {\calC^{V}_{200}}}}{4f^2_\pi} \left\{ \left< {N^{\dagger}}N \right> \left< {N^{\dagger}}N \right> 
+ 4\left< {N^{\dagger}}t_3 N \right> \left< {N^{\dagger}}t_3 N \right> 
\right\}
\ea\ee
with only {\it four} independent chiral coefficients:
\be\ba
\label{Nuclear_Coefficients}
C^{S}_{200}=&C^{T=0}_{S} \\
-{\overline {C^{S}_{200}}} =& \tfrac{1}{4} \left[C^{T=0}_{S}+ 5 C^{T=1}_{S} \right. \\
&+6 \left( C^{T=0}_{T} + C^{T=1}_{T}\right) +\left. \left( C^{T=0}_{P}+  C^{T=1}_{P} \right) \right], \\
C^{V}_{200}=&C^{T=0}_{V}, \\
-{\overline {C^{V}_{200}}} =& \shalf \left[ -C^{T=0}_{V}+C^{T=0}_{A} + C^{T=1}_{V}+ C^{T=1}_{A}\right].
\ea\ee
To simplify the notation and to retain the connection with previous work \cite{Chin197424} we introduce:
\be
\calC_V^2\equiv 
\frac{1}{f_\pi^2}
\left(C^V_{200}
   -\frac{1}{2}\overline{C^V_{200}}\right), \\
\label{Walecka_parameters_V}
\ee
\be
\calC_S^2\equiv 
-\frac{1}{f_\pi^2}
	\left(C^S_{200}
	-\frac{1}{2}\overline{C^S_{200}}\right) \,.
\label{Walecka_parameters_S}
\ee
For brevity, we also define:
\be
\overline{\calC_V^2} \equiv 
\frac{1}{f_\pi^2}
	\overline{C^V_{200}},\\
\label{Exchange_parameters_V}
\ee
\be
\overline{\calC_S^2} \equiv 
\frac{1}{f_\pi^2} \overline{C^S_{200}}.
\label{Exchange_parameters_S}
\ee
In \eqref{LagrangianMF-StaticXNL} the operator $\Theta$ is given by:
\be
\Theta \equiv 
-m^N - \widehat{C_{200}^S}
- \widehat{C_{200}^V}\gamma^0\,, 
\ee
with:
\be\ba
  \widehat{C_{200}^V}&\equiv 
C^2_V \, \left<{N^\dagger}N\right>
 -2\, {\overline{\calC^2_V}} \left<{N^\dagger}t_3 N\right>t_3,  \\
 \widehat{C_{200}^S}&\equiv -\calC^2_S \, \left<{\overline N}N\right> 
-2\, {\overline{\calC^2_S}} \left<{\overline N}t_3 N\right>t_3,  \\
 0 &= \left[t_3\, ,   \widehat{C_{200}^S}\right] =  \left[t_3\, ,   \widehat{C_{200}^V}\gamma^0\right]  = \left[t_3\, , \Theta \right].
\ea\ee
We have ignored possible excited states that contribute to fluctuations in the nuclear density and which are beyond the scope of this paper.

The \SchiNLb Lagrangian offers a significant improvement in the predictive power of the theory,
while still providing sufficient free parameters 
to balance vector repulsive forces against scalar attractive forces
when fitting 
(to order $\Lambda_{\chi SB}^0$) 
non-topological-soliton and Skyrme nuclear models 
to the experimentally observed structure of ground state nuclei.  
Further simplification results for a sufficiently large number of nucleons: simple Hartree analysis of \eqref{BosonExchangeContactInteractions} is equivalent to more accurate Hartree-Fock analysis of the same Lagrangian without spinor-interchange terms.

We now see that, inside the \SchiNL, 
a nucleon living in the self-consistent field of the other nucleons obeys 
the Dirac equation 
\begin{eqnarray}
\label{Dirac_equation}
0&=&\left( i \gamma^\mu \, \partial_\mu 
	+\Theta\right) N\,.
\end{eqnarray}
Baryon-number and the third component of isospin are both conserved; i.e., the associated currents $J^\mu_{Baryon}\equiv{\overline N}\gamma^\mu N$
and $J^\mu_{3}\equiv{\overline N}\gamma^\mu t_3 N$ are both divergence-free.
The neutral axial-vector current 
$J^{5,\mu}_{8}\equiv \frac{\sqrt{3}}{2}{\overline N}\gamma^\mu \gamma^5 N$,
corresponding to the projection onto SU(2)
of the NGB $\eta$ particle, part of the unbroken SU$(3)_L\times$SU$(3)_R$ meson octet,
is also divergence free,
\be\ba
\frac{2}{\sqrt{3}}\left< i\partial_\mu J^{5,\mu}_{8}\right>
&=\left< {\overline N} \left\{  \Theta,\gamma^5  \right\} N\right>
\\
&=2\left< {\overline N} \left( -m^N - \widehat{C_{200}^S} \right) \gamma^5 N\right>
\\
&\simeq 0.
\ea\ee
This result can be understood as a statement that
the $\eta$ particle 
cannot survive in the parity-even interior of a \SchiNL,
since it is a NGB pseudo-scalar in the chiral limit.  
Similarly, the 3rd component of the axial vector current  
is divergence-free; i.e.,
\be\ba
\left< i\partial_\mu J^{5,\mu}_{3}\right> 
&=\left< {\overline N} \left\{  \Theta,\gamma^5  \right\} t_3 N\right>
 \\
 &=2\left< {\overline N} \left( -m^N - \widehat{C_{200}^S}\right) \gamma^5 t_3 N\right>  \\
&\simeq 0,
\ea\ee
because the SU$(2)\chi$PT $\pi_3$ particle 
is also a NGB pseudo-scalar in the chiral limit,
and cannot survive in the interior of a parity-even \SchiNL.

Even though explicit pion and eta fields
vanish in \SchiNL, 
their quantum numbers reappear in its PCAC properties 
from nucleon bi-linears and four-nucleon terms in 
the divergences of axial vector currents.
That these average to zero in \SchiNLb plays a crucial role in the conservation of axial-vector currents within the liquid.

It is now straightforward to see that, in the liquid approximation, 
a homogeneous \chiPTtwb nucleon liquid drop with no meson condensate 
satisfies all relevant CVC and PCAC equations. 
As shown in Appendix \ref{Nuclear_current_section}, most of the space-time components of 
the three SU(2)$_{L+R}$ vector currents $J^\mu_a$ 
and three axial vector currents $J^{5\mu}_a$ vanish: only $J^0_3$ is nonzero in \SchiNL.

The neutral SU$(3)_L\times$SU$(3)_R$ currents are conserved $\left< \partial_\mu J^{\mu}_8 \right>=0$ and $\left< \partial_\mu J^{5;\mu}_8\right>= 0$
 in the \SchiNLb mean field.  In addition, the neutral SU(3)$_{L+R}$ vector current's spatial components $J^{\mu =1,2,3}_8$ 
and the axial-vector currents $J^{5;\mu}_8$
all vanish.  Only  $J^{0}_8$, proportional to the baryon number density, survives  in the \SchiNLb  mean field.
 
Since \SchiNLb chiral nuclear liquids 
satisfy all relevant $\chi PT$ CVC and PCAC equations 
in the liquid phase, 
they are true solutions of the all-orders-renormalized tree-level semi-classical liquid equations of motion truncated at ${\mathcal O}(\Lambda_{\chi SB}^0 )$. 

\section{Nuclei and neutron stars as mean-field static $\chi$NL}

\subsection{Thomas-Fermi non-topological solitons, liquid drops and the semi-empirical mass formula}
\label{Nontopological_Solitons}

Mean-field \SchiNLb non-topological solitons are solutions of \chiPT semi-classical symmetries, obeying all CVC and PCAC conservation laws. 
They have zero internal and external pressure. 
The nuclear liquid-drop model and 
Bethe-von Weizs\"acker SEMF emerge -- 
with correct nuclear density, 
and saturation and asymmetry energies -- 
in an explicit Thomas-Fermi construction. 

In Appendix \ref{Thomas_Fermi}, we construct explicit liquid 
mean field Static$\chi NL$ solutions based on \eqref{LagrangianMF-StaticXNL}, 
constrained to order 
$4\pi f_\pi  \approx \Lambda_{\chi SB} \simeq 1$ GeV 
and $\Lambda^0_{\chi SB}$ naive power-counting,
in an independent-nucleon model, using the Thomas-Fermi free-particle approximation.\footnote{An effective Lagrangian, built from ${\cal O}(\Lambda_{\chi SB})$ free nucleons and ${\cal O}(\Lambda_{\chi SB}^0)$ point-coupling interaction-operators, was also identified by G. Gelmini and B. Ritzi \cite{Gelmini1995.431}.  However, it does not correspond to Chin-Walecka infinite symmetric $Z=N$ nuclear matter, and the authors constructed no $Z=N$ bound-state non-topological-solitons with zero internal and external pressure, which could therefore survive in an external vacuum.}
	
Constant-density non-topological solitons, 
i.e., liquid drops comprised entirely of nucleons, 
emerge as homogeneous and isotropic 
semi-classical static solutions 
with internal and external pressures both zero.
Their surface is a step function. 
Ignoring electro-magnetism, 
nuclear matter and finite nuclei 
then have identical microscopic structure, 
serving as a model of the ground state
of both  infinite nuclear matter
and finite liquid drops.
There is no need for an additional confining interaction
to define the finite-drop surface.
With even proton number $Z$, and even neutron number $N$, 
nucleons are arranged in pairs 
so that local expectation values for spin vanish, $< {\vec s}> \simeq 0$. 
The microscopic structure is also spherically symmetric,
so that local momenta have a vanishing expectation value, $<{\vec k}>\simeq 0$.
Consequently, 
total spin ${\vec S}= 0$ 
and total momentum ${\vec K} = 0$ in the center-of-mass.

The semi-empirical mass formula \cite{vonWeizsacker1935,Rohlf1994} is:
\be\ba
\label{Semi_Empirical_Mass_Formula}
M(Z,N)&=Z m_p+N m_n-E_{B}, \\
E_{B} &= E_{B}^{Vol} + E_{B}^{Surf} + E_{B}^{Pair},\\
E_{B}^{Vol}/A &\equiv a_{V} -a_{Asym}\,X^2
 -a_{C}\frac{Z(Z-1)}{A^{4/3}}, \\
E_{B}^{Surf}/A &\equiv -\frac{a_{S}}{A^{1/3}}, \\
E_{B}^{Pair}/A &\equiv a_{Pair}\,\frac{\delta_0(Z,N)}{A^{3/2}}.
\ea
\ee
with $A=Z+N$; $X$ is the neutron excess:
\be
X \equiv \biggl(\frac{N-Z}{N+Z}\biggr),
\ee
and
\begin{equation}
	\delta_0 \equiv \begin{cases}  
	+1\,& \quad\mathrm{for}\; Z\; \mathrm{even},~N~\mathrm{even}\,,\\
	-1\,& \quad\mathrm{for}\;Z\;\mathrm{odd},~N~\mathrm{odd}\,,\\
	0\, & \quad\mathrm{for}\;$A=Z+N$\;\mathrm{odd}\,.
	\end{cases}
\end{equation}
From  \cite{Rohlf1994} we use:  $a_{V}=15.75$ MeV, 
$a_{S}=17.8$ MeV, 
$a_{C}= 0.711$ MeV, 
$a_{Asym}=23.7$ MeV, and	
$a_{Pair}=11.18$ MeV.

We show in Appendix D that the SEMF is (almost) an \SUtwLR $\chi$PT non-topological soliton prediction.   We first display symmetric $Z = N$ ground state	zero-pressure Hartree-Fock non-topological-soliton solutions,
fit to inferred experimental values	for symmetric-nuclear-matter density and volume binding energy and find:
\begin{equation}
\begin{aligned}
\label{ZEqualNFit}
\calC_V^2
&\,= \,1.893\,\frac{1}{f_\pi^2},  \\
\calC_S^2
&\,=\, 2.580\,\frac{1}{f_\pi^2}.
\end{aligned}
\end{equation}
These values were obtained by fitting to a Fermi momentum ($k_{Fermi} =1.42/\mathrm{fm}$) 
and saturated volume energy ($E_{binding}/nucleon = 15.75$MeV).
We observe that for heavy nuclei $X^2 \ll 1$, and work to leading order in that small quantity. 
In Appendix \ref{ZneqNappendix}, we derive asymmetric $Z \neq N$ nuclear matter, 
for which fermion-exchange terms are crucial, 
fitting to $a_{Asym}=23.7$ MeV.
For $k_F = 1.42/fm$ we find:
\begin{eqnarray}
\label{Asymmetry}
{\overline{\calC^2_V}}  = 0.61 \,\,\frac{1}{f_\pi^2}.
\end{eqnarray}
Additional results for $\overline{C_{200}^V}$ are given in Appendix \ref{CV200BAR}.  
Combining \eqref{ZEqualNFit} and \eqref{Asymmetry} using \eqref{Walecka_parameters_V} and \eqref{Exchange_parameters_V} gives:
\begin{equation}
\begin{aligned}
\label{CV200_final_value}
C^V_{200} = 2.198 \,.
\end{aligned}
\end{equation}

In practice, there is very little sensitivity to our 4th independent chiral coefficient $\overline{C^S_{200}}$:
this in agreement with
Niksic \cite{Niksic2008.034318} et al., who argue that, although the total iso-vector strength has a relatively well-defined value, 
the distribution between 
the iso-vector Lorentz-scalar $\vec \delta$ exchange channel, and the iso-vector Lorentz-vector ${\vec \rho}_\mu$ exchange  channel,
is not determined by ground state data.  We have assumed 
$(\frac{Z-N}{Z+N})^2 \ll 1$.  In addition, we have
\begin{equation}
\begin{aligned}
\bran N^\dagger N \ketn - \bran {\bar N} N \ketn &\simeq \frac{3}{10}\frac{k_F^2}{m^2_{*8}}\bran N^\dagger N \ketn\\
&= (0.0762)\, \bran N^\dagger N\ketn << \bran N^\dagger N \ketn
\label{Non_relativistic_densities}
\end{aligned}
\end{equation}
where $k_F = 279.7$ MeV and where $m_{*8} \equiv \half (m_{*p}+m_{*n}) = 555$ MeV which follows from the Thomas Fermi solution in as found in Appendix \ref{ZEqualNappendix}.  It follows that only the combination 
 $\big( \overline{C^V_{200}} + \overline{C^S_{200}} \big) $ 
 can strictly be fit to our ${\cal O}(\Lambda_{\chi SB}^0)$ Static$\chi NL$ accuracy.  Therefore, for convenience and without loss of generality, we choose 
\begin{eqnarray}
\label{CSBar200}
\overline{C^S_{200}} &=&0\,.
\end{eqnarray}
All coefficients in \eqref{ZEqualNFit}, \eqref{Asymmetry}, \eqref{CV200_final_value}, and \eqref{CSBar200} then obey naive $\sim{\cal O}(1)$ dimensional power counting, and so are legitimate natural chiral coefficients. Note the fine-tuning between 
${C^V_{200}}=2.198$ and ${C^S_{200}}=-2.580$ 
in \eqref{ZEqualNFit} and \eqref{CV200_final_value} inherited from R. Serber's and J.D. Walecka's 1974 quadratic models \cite{Chin197424}, \cite{Serber1974up} and \cite{Serber1976718}. That fine-tuning is alleviated in \eqref{ZEqualNFit} by the our inclusion of 
${\vec \rho}_\mu$ exchange, necessary to Static$\chi NL$s.
Equations \eqref{ZEqualNFit}, \eqref{Asymmetry}, \eqref{CV200_final_value} and \eqref{CSBar200}
all satisfy naive dimensional power-counting ${\cal O}(1)$ naturalness, and so are legitimate chiral coefficients.  The astute reader will notice that the difference 
\eqref{Non_relativistic_densities}
is of the same order as the next terms in the chiral expansion.  Although we have calculated self-consistently in powers of $\Lambda_{\chi SB}$ in chiral perturbation theory, terms of order $\Lambda_{\chi SB}^{-1}$ must still play an important role
in the nontopological soliton solutions. Indeed, it is inconsistent to neglect them.  We hope to return to this question in future work.

The SEMF is closely associated 
with Gamow's nuclear liquid-drop model (NLDM).
Recall that, 
following Walecka's infinite symmetric nuclear matter 
(and neutron matter), 
we have imposed on the Thomas-Fermi mean field 
the condition that the pressure vanish 
both internally and externally, 
not only at the surface of a finite ``drop.''
Our non-topological soliton nuclei 
therefore resemble ice cream balls 
scooped from an infinite vat \cite{Boguta1983.34}, 
more than they do conventional liquid drops. 

We clearly have no right to use the Thomas-Fermi approximation
to calculate the surface and pairing energies, $E_{B}^{Surf}$ and $E_{B}^{Pair}$ of \eqref{Semi_Empirical_Mass_Formula},
at order $ \Lambda_{\chi SB}$ and $\Lambda^0_{\chi SB}$ 
in the spontaneously broken theory. 
Unsurprisingly, 
the surface energy calculated entirely as a change in density gives incorrect $a_{S}$.
However, there exist ${\mathcal O}\bigl(\Lambda^{-2}_{\chi SB}\bigr)$  
nuclear-surface \chiPTtwb terms
that might replace the scalar $\sigma$ particle in the Chin-Walecka model 
in describing the nuclear surface \cite{Chin197424, Serr197810, Serot19861}, namely 
\be
{\mathcal L}_{\chi PT}^{Surf} = -\tfrac{1}{2}\frac{C_{220}}{\Lambda_{\chi_{SB}}^{2}}\,\partial_\nu \left({\bar N} N\right) \partial^\nu \left({\bar N} N \right),
\ee
with an ${\mathcal O}(1)$ constant $C_{220}$,
obeying naturalness.
${\mathcal L}_{\chi PT}^{Surf}$ 
is invariant under non-linear \SUtwLR transformations including pions, 
but is automatically pionless, even without the liquid approximation. 
It contains no dangerous $\partial_0 \sim m_N$ nucleon mass terms, 
so non-relativistic re-ordering is unnecessary. 
Nucleon-exchange and spinor-interchange interactions must also be included.

Meanwhile, calculation of $a_{Pair}$ involves understanding low-level excited states, 
such as Z-odd N-odd states which we have ignored in our study of the Lagrangian \eqref{BosonExchangeContactInteractions}, which are beyond the scope of this paper, and will likely require explicit pions 
lying outside semi-classical pionless \chiPTtw. 

\subsection{Neutron Stars}
Putting aside exotica (i.e., quark condensates, pion condensates, strange-kaon condensates, etc.), 
we conjecture that much of the structure of neutron stars may be traced directly to 2-massless-quark QCD, and thus directly to the Standard Model. 
This will be explored further in a companion paper. 
Here we note only that the models of Harrison \& Wheeler \cite{Harrison1958}, Salpeter \cite{Salpeter1961669} and Baym, Pethic and Sutherland \cite{Baym1971299}, 
are all based on the Bethe-von Weizs\"acker semi-empirical mass formula
\cite{Shapiro1983}.
They would therefore seem to follow from Static$\chi NL$;
however, we do not yet know how well the observed chart of nuclides and these neutron-star models match
the "ice-cream scoop" Static$\chi NL$ no-surface SEMF, augmented by Coulomb repulsion; 
i.e., \eqref{Semi_Empirical_Mass_Formula} with 
$E_{B}^{Pair}$ set to zero.

\subsection{Shell structure from
chiral symmetry breaking?}

We conjecture here that non-topological Static $\chi NL$ solitons could, 
with inclusion of explicit {axial symmetry} breaking, 
be re-quantized to incorporate 
no-core nuclear shell structure and magic numbers, 
as imagined in \cite{Lynn1993281}.
Lynn first introduced the idea \cite{Lynn1993281}
that \SUtwLR $\chi$PT could admit a liquid phase.
Like ours, his Lagrangian included only terms of ${\mathcal O}(\Lambda_{\chi SB})$ and ${\mathcal O}(\Lambda_{\chi SB}^0)$.
Though he did not anticipate Static$\chi NL$s, 
he was careful to include only and all those terms that respect the SU$(2)_L\times$SU$(2)_R$ semi-classical symmetries - i.e., of quantum nucleons and classical pions - discussed in this paper. 
These included strong interaction terms that survive the chiral limit, as well as explicit axial breaking terms that do not. 

The purpose of \cite{Lynn1993281} was to generate a "no-core" classical static spherical  central potential for $\vert {\vec\pi} \vert$, 
in which all of the quantum nucleons moved, and thus plausibly shell structure for certain heavy even-even ground state spin-zero spherical nuclei. 
It now seems advantageous to focus on doubly-magic or spherically-magic nuclides.

Such shell structure is plausible in semi-classical \SUtwLR $\chi$PT because
the explicit symmetry-breaking terms have naive operator power counting $m\!=\!0$, $l\!=\!1$, $n\!=\!1$ in
\eqref{L_chiPT_full}. Ignoring $\pi^\pm-\pi^0$ mass splitting, these are
\be\ba
\label{Symmetry_Breaking}
\bran L_{\chi PT}^{N;\chi SB}\ketn &\simeq \overline{m}\left(a_1+2 a_3\right)  
 \left[1-\cos{\frac{\vert {\vec\pi} \vert}{f_\pi}} \right] \, {\big<\bar N} N \big>, \\
&\equiv \beta \sigma_{\pi N} \left[1-\cos{\frac{\vert {\vec\pi} \vert}{f_\pi}} \right] {\big< \bar N} N \big>,
\ea\ee
with
\be
\overline{m} \equiv \tfrac{1}{2}(m_{u} + m_{d}),
\ee
and with experimental parameters:
\be\ba
(a_1,a_2,a_3) &= (0.28, -0.56, 1.3\pm 0.2)\\
(m_{u},\,m_{d},\,\sigma_{\pi N}) &= (6,\,12,\,60) \,\mathrm{MeV}\\
\beta &= 0.864\pm0.120
\ea\ee
measured in SU$(3)_L\times$SU$(3)_R$ $\chi PT$ processes 
\cite{Georgi1984} and \cite{Lynn2010}.

Since $\bran L_{\chi PT}^{N;\chi SB} \ketn > 0$, the explicit symmetry-breaking terms lower the effective nucleon mass inside a static $\pi = \vert {\vec\pi} \vert$ condensate.

We conjecture that semi-classical \SUtwLR $\chi$PT (i.e., including all ${\mathcal O}(\Lambda_{\chi SB})$ and ${\mathcal O}(\Lambda_{\chi SB}^0)$ non-strange analytic naive operator power-counting terms, both those from the chiral-limit and those from explicit $m_{u},m_{d}\neq 0$ chiral symmetry breaking) applied to certain {\it finite} nuclei, nuclear and neutron matter and neutron stars will give a reasonable match to their structure.

\section{Relation of Static$\chi NL$ to pionless EFT}
\label{Pionless_EFT_vs_Static_Chiral_NL}
Our pionless static chiral nuclear liquid solution bears superficial resemblance to results from pionless EFT \cite{Hammer2020.025004}: both are ``pionless.''  They are both pionless for different reasons, however.  	Pionless EFTS are pionless because the pions have been ``integrated out'' and so are valid for momenta less than the pion mass.  Static$\chi NL$ are pionless because the pionic source terms vanish in even-even, spin-zero spherical nuclei: here we work in the chiral limit of vanishing pion mass.  The soliton solution has $k_F\simeq 280$ MeV.  In fitting the parameters $C_{\cal A}$ in Eq. (2),
we must fit to inferred infinite-nuclear-matter data.  As pointed out by Hammer et al. \cite{Hammer2020.025004} perturbation theory cannot be used to relate the coupling constants in the two theories.  In future work one might hope to relate the coupling constants of Static$\chi NL$ to those of pionless and of Halo/Cluster EFTs \cite{Hammer2020.025004}.

U. van Kolck and the Pionless EFT community like to reveal relationships among their results by plotting them on the complex $Re (k)-Im (k)$ momentum plane inside the circle $\vert k\vert \leq \Lpionless ^A < m_\pi$. To that disc we add an orthogonal $A=Z+N$ axis -- forming a 3-D cylindrical $ Re (k)-Im (k)-A$ volume    
-- and highlight some Pionless EFT results. 
In the $A=2$ plane, $N-N$ elastic scattering is properly compared to Nijmegen data and lies along positive $Re(k)$. 
The $-2.2$ MeV bound deuteron is at $k_{Pole}^{^3S_1}$ on the positive $Im(k)$ axis, 
while the shallow resonance is at $k_{Pole}^{^1S_0}$ on the negative $Im(k)$ axis. 
The $A=4$ plane places the deeply bound ($-28.296$ Mev) $\alpha$ particle $(\sim \, ^4_2He_2)$ at positive $Im(k)$.

Halo/cluster EFT at $A \geq 5$ has no pions, and is mathematically similar to Pionless EFT, becoming Pionless EFT for light nuclei when the cores are nucleons.  
We plot only the classic example $^6_2He_4$, where the energy required to remove the cluster ($\alpha$ particle), or either of the two halo nucleons, is much less than to break up the cluster. 
It lies on the $A=6$ plane at positive $Im (k)$.

In order to plot our Thomas-Fermi Static$\chi NL$ results from Appendix \ref{Thomas_Fermi} and show their position relative to Pionless EFT, we add an annulus to that Pionless EFT cylinder, extending the radius of its $Re(k) - Im(k)$ base to the region $\Lpionless ^A < \vert k\vert \leq \Lambda_{Static \chi NL}^A = \Lambda_{\chi SB} \sim 1$ GeV.  
Our bound-state Static$\chi NL$ "ice-cream-scoop" nuclei are then horizontal lines along $Im(k)$, 
in the positive $Im(k) - A$ quarter-plane,
with $0 \leq \vert k\vert \leq k_F\simeq 280$ MeV: they intersect the $A$ axis at $A_{\mathrm{Even}\mathrm{Even}}=Z_{\mathrm{Even}}+N_{\mathrm{Even}} \geq 4$.
For visual simplicity, 
we plot symmetric $Z=N$ Static$\chi NL$ nuclei 
only for $^{28}_{14}$Si and  $^{40}_{20}$Ca, 
at $A=28, 40$. 
We show asymmetric Static$\chi NL$ only for $^{48}_{20}$Ca, $^{60}_{28}$Ni, $^{90}_{40}$Zr and $^{208}_{\,\,82}$Pb with $X^2\ll1$.  
For further pedagogical simplicity, 
we have averaged 
$\half(k_F^p + k_F^n)\approx k_F \simeq 280$ MeV. 

Going forward,
an important challenge is to find an \SUtwLR$\,\chi PT$ integration
of the physics of Static$\chi NL$ and that of Pionless EFT and halo/cluster EFT.

In the Summary of the 1985 Paris Conference on Nuclear Physics with Electro-magnetic Probes,
Torleif Ericson  \cite{Ericson1985.507} showed just how many facets there are to the nuclear ``truth''  -- different physical domains require different descriptions, each of which is the truth for that domain.
If Static $\chi$NL as derived from the symmetries of QCD describes heavy (spin-zero even-even) spherical nuclei, 
its truth may be difficult to relate directly to accurate descriptions of other physical domains.  
\begin{figure}[htp]
\centering
        \includegraphics[width=3.375 in]{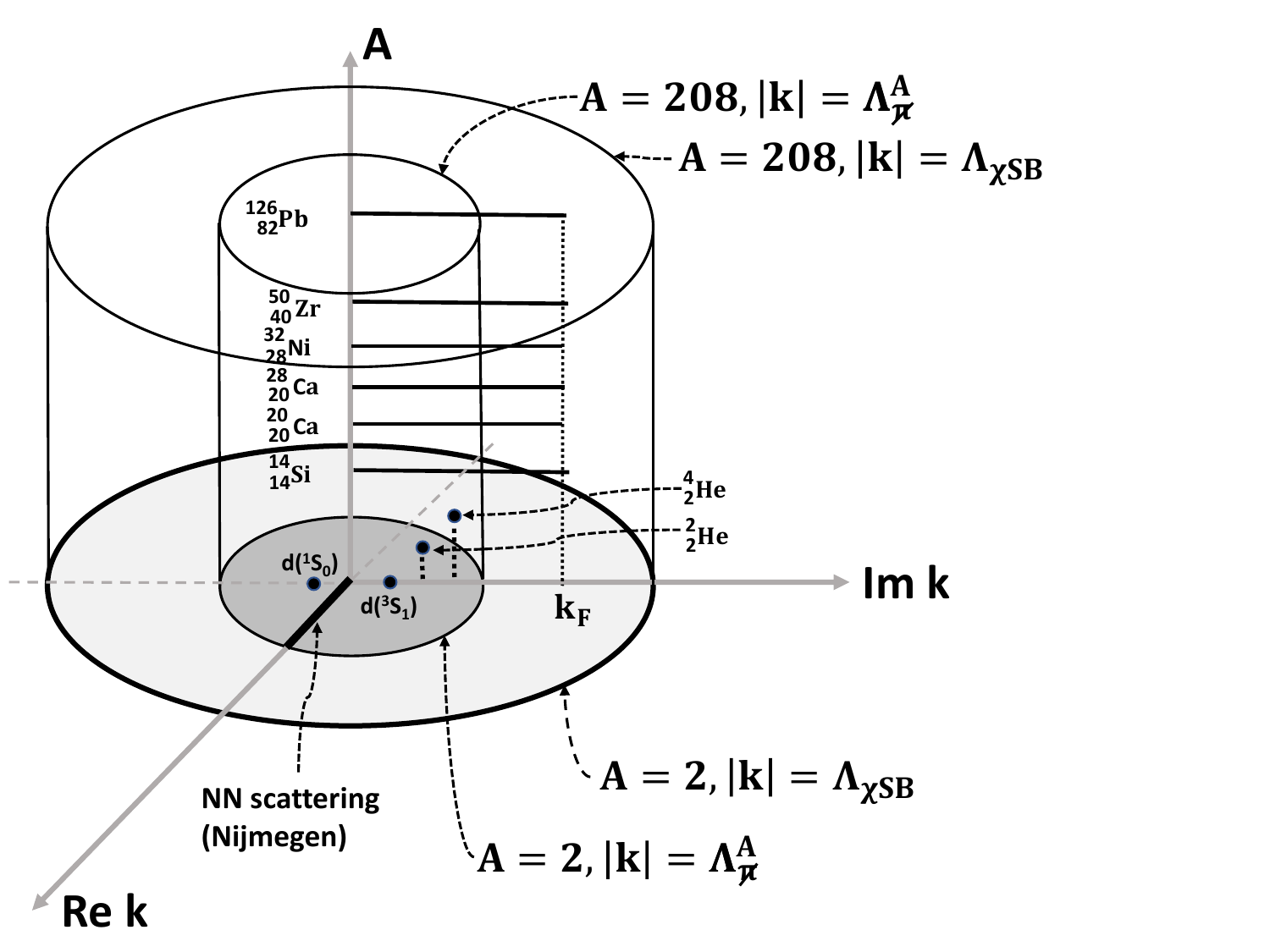}
        \caption{
        Illustration (not to scale) of the domains 
        of applicability of various analytic treatments of nuclear systems plotted in the 3-dimensional space 
        defined by complex momentum $(\text{Re} k, \text{Im} k)$
        and atomic/baryon number $A$.  
        At the base sits the $A=2$ complex $k$ plane.
        Pionless effective field theory is valid
        inside the cylinder whose base is the disk
        with radius $\Lpionless^A< m_\pi$.
        The even-even spin-zero nuclei to which
        the chiral nuclear liquid treatment of this paper
        are applicable are shown here: 
        $^{28}_{14}$Si, 
        $^{40}_{20}$Ca,
        $^{48}_{20}$Ca,
        $^{60}_{28}$Ni,
        $^{90}_{40}$Zr, and
        $^{208}_{\,\,82}$Pb.
        Their treatment incorporates $k$'s along 
        the $\text{Im} k$ axis from $0$ 
        to $k_{\mathrm Fermi}\ll \Lambda_{\chi SB}$.       
				See text for further details.
        \label{fig:Bira}
        }
\end{figure}
\section{Conclusions}
In this paper, we have explored heavy symmetric nuclei in a semi-classical approach starting with Chiral EFT that respects the global symmetries of QCD.
In this, we have been 
guided by two key observations: 
that nuclei are made of protons and neutrons,
not quarks;
and that the up and down quarks, 
which are the fermionic constituents of the protons and neutrons,
are much lighter than the principal mass scales of QCD, 
such as the proton and neutron masses.
Taken together, 
these strongly suggest that the full complexity of the Standard Model
can largely be captured, for the purposes of nuclear physics, 
by an effective field theory (EFT) --
\SUtwLR chiral perturbation theory (\chiPTtw)
of protons and neutrons.

Building on this longstanding insight, 
we have studied the chiral limit of spontaneously broken SU$(2)_L\times$SU$(2)_R$ (i.e., \chiPTtw), 
including only operators of order 
$\Lambda_{\chi SB}$ and $\Lambda_{\chi SB}^{0}$.
We find that
\chiPTtwb of protons,
neutrons and three pseudo Nambu-Goldstone Bosons pions - 
admits a semi-classical liquid phase,
a Static Chiral Nucleon Liquid (\SchiNL).
\SchiNL s are made entirely of nucleons,
with approximately zero anti-proton and anti-neutron content.
They are parity even and time-independent.
As we have studied them so far, 
not just the total nuclear spin ${\vec S}= 0$, 
but also the local expectation value for spin $< {\vec s}> \simeq 0$.
Similarly, 
the nucleon momenta vanish locally in the spherically symmetric \SchiNLb rest frame.
For these reasons, 
our study of \SchiNLb is applicable 
to bulk ground state spin-zero nuclear matter,
and to the ground state of appropriate spin-zero parity-even nuclei 
with an even number $Z$ of protons and an even number $N$ of neutrons.

We classify these solutions of \chiPTtwb as ``liquid" because
energy is required both to pull the constituent nucleons further apart
and to push them closer together.
This is analogous with the balancing of the
attractive Lorentz-scalar $\sigma$-exchange force 
and 
the repulsive Lorentz-vector $\omega_\mu$-exchange force 
in the Walecka model.
The nucleon number density therefore takes a saturated value
even in zero external pressure (e.g. in the absence of gravity),
so the material is not a ``gas."
Meanwhile they are statistically homogeneous and isotropic,
lacking the reduced symmetries of crystals or other solids.

We have shown that in this ground state liquid phase, 
the expectation values 
of many of the allowed operators 
of the most general \chiPTtwb Lagrangian
vanish or are small.
Going forward, 
it is imperative to understand the effects of 
of excited nucleon states to the spectra of heavy nuclei.  

We have also shown that this spontaneously broken ground state liquid phase 
does not support a classical pion field --
infrared pions decouple from this solution.
We expect that this emergence of ``semi-classical pionless \chiPTtw''
is at the heart of the apparent theoretical independence 
of much successful nuclear structure physics from pion properties 
such as the pion mass.

We have constructed explicit \SchiNL's in the Thomas-Fermi approximation,
demonstrating the existence of zero-pressure non-topological soliton \SchiNLb solutions 
with macroscopic (infinite nuclear matter) 
and microscopic (heavy nuclear ground states). 

\section*{Acknowledgments}
\indent
We thank David M. Jacobs for his early contributions to this work.
BJC, BWL and GDS thank Bira van Kolck for crucial discussions.  We would also like to thank the referee for suggestions that contributed significantly to the quality of the paper.
BJC and BWL thank IPNO, Universite de Paris Sud, for their hospitality in summer 2019; while BJC thanks CWRU for its hospitality in fall 2019.  GDS is partially supported by a grant from the US DOE to the particle-astrophysics theory group at CWRU.

\bibliographystyle{h-physrev}
\bibliography{../bib/ChiralNuclearLiquids}
\appendix
\numberwithin{equation}{section}
\section{\SUtwLR $\chi$ PT of a nucleon doublet
and a pion
triplet in the spontaneously broken (i.e., chiral) limit}
\label{app:chiptLagrangian}

The chiral symmetry of two light quark flavors in QCD, 
together with the symmetry-breaking and Goldstone's theorem, 
makes it possible to obtain an approximate solution to QCD at low energies 
using a \SUtwLR EFT,
where the degrees of freedom are hadrons \cite{
	Weinberg19681568, Weinberg1978.327, Coleman19692239, Callan19692247, 
	Gasser1984142, Gasser1985465, Manohar1984189, Georgi1993187, Georgi1984}.   
In particular, the non-linear \chiPTtwb effective Lagrangian 
has been shown to successfully model the interactions of pions with nucleons,
where a perturbation expansion 
(e.g., in soft momentum $|\vec{k}|/\Lambda_{\chi SB} \ll 1$,  
baryon number density $\frac{<N^\dagger N>}{f^2_\pi \Lambda_{\chi SB}} \ll 1$, 
for chiral symmetry breaking scale $\Lambda_{\chi SB} \approx 1$ GeV) 
has demonstrated predictive power. 
Such $naive$ power-counting in $\Lambda_{\chi SB}^{-1}$ 
includes all analytic quantum-loop effects 
into experimentally measurable coefficients of \SUtwLR 
current-algebraic operators 
obedient to the global symmetries of QCD, 
with light-quark masses generating additional explicit 
chiral-symmetry-breaking terms. 
Therefore, \SUtwLR $\chi$PT tree-level calculations 
with a naive power-counting effective Lagrangian 
are to be regarded as true predictions of QCD 
and the Standard Model 
of elementary particles. 

\subsection{Non-linear transformation properties}
\label{NonLinearTransformationProperties}

We present the Lagrangian of \SUtwLR$\chi$PT
of a nucleon doublet and a pNGB triplet. 
We employ the defining SU(2) strong isospin representation of
unitary $2\stimes2$ Pauli matrices $\tau_a$, 
with asymmetric structure constants 
$f_{abc}=\epsilon_{abc}$
\begin{equation}
  \begin{aligned}
 	\label{SU_three_algebra}
    t_a = \frac{\tau_a}{2}\,, \quad a = 1,3 \\
    \Tr(t_a t_b) = \frac{\delta_{ab}}{2} \\
    \left [ t_a, t_b \right] = if_{abc}t_c \\
    \left \{t_a, t_b \right\} = \frac{\delta_{ab}}{2}     \,.
  \end{aligned}
\end{equation} 
The vector and axial-vector charges obey the algebra:
\begin{equation}
\begin{aligned}
\label{VandA_charges_algebra}
\left [ Q^{L+R}_{a}, Q^{L+R}_{b} \right] &=& i f_{abc}Q^{L+R}_c \\
\left [ Q^{L-R}_{a}, Q^{L-R}_{b} \right] &=& i f_{abc}Q^{L+R}_c \\
\left [ Q^{L+R}_{a}, Q^{L-R}_{b} \right] &=& i f_{abc}Q^{L-R}_c \,.
\end{aligned}
\end{equation} 

We consider a triplet representation of NGBs, 
\begin{equation}
  \begin{aligned}
 \label{pGoctetrepn1}
    \pi_at_a = \frac{1}{\sqrt{2}}\left [
    \begin{tabular}{ccc} 
    $\frac{\pi^0}{\sqrt{2}}$ & $\pi^+$ \\ 
    $\pi^-$ & $-\frac{\pi^0}{\sqrt{2}}$      \\
    \end{tabular} \right]
  \end{aligned}
\end{equation}
and a doublet of nucleons,
\begin{equation}
  \begin{aligned}
  \label{Boctetrepn}
    N = \left [
    \begin{tabular}{ccc} 
     $p$ \\ 
     $n$ \\
    \end{tabular} \right].
  \end{aligned}
\end{equation}

For pedagogical simplicity, 
representations of higher mass are neglected, 
even though the \SUtwLR baryon decuplet 
(especially $\Delta_{1232}$) 
is known to have important nuclear structure \cite{Ericson1988} 
and scattering \cite{Jenkins2002242001} effects.

Since \chiPTtwb matrix elements 
are independent of representation \cite{Coleman19692239, Callan19692247},
we choose a representation \cite{Manohar1984189, Georgi1993187, Georgi1984} 
where
  the NGB triplet has only derivative couplings, 
\begin{equation}
\label{Sigmarepn}
\Sigma \equiv \expon(2i \pi_a \frac{t_a}{f_\pi})\,.
\end{equation}
Under a unitary global 
\SUtwLR
transformation, 
given by $L\equiv \exp(il_at_a)$ and $R \equiv \exp(ir_at_a)$,
\begin{equation}
\label{Sigmatransfn}
\Sigma \rightarrow \Sigma' = L\Sigma R^\dagger\,.
\end{equation}
It also proves useful to introduce the ``square root'' of $\Sigma$
\begin{equation}
  \xi \equiv  \expon(i \pi_a \frac{t_a}{f_\pi}),
\end{equation}
which transforms as
\begin{eqnarray}
    \xi &\rightarrow& \xi^\prime =  \expon(i\pi^\prime_a\frac{t_a}{f_\pi})\,.
\end{eqnarray}
We observe that
\begin{eqnarray}
    \xi^\prime &=& L\xi U^\dagger = U\xi R^\dagger,
\end{eqnarray} 
for a certain unitary local transformation matrix $U(L,R,\pi_a(t,x))$.

The vector and axial-vector NGB currents
\begin{equation}
\begin{aligned}
V_\mu \equiv \frac{1}{2}(\xi^\dagger\partial_\mu\xi + \xi\partial_\mu\xi^\dagger),\\
A_\mu \equiv \frac{i}{2}(\xi^\dagger \partial_\mu \xi - \xi\partial_\mu\xi^\dagger),
\end{aligned}
\end{equation}
transform straightforwardly as:
\begin{equation}
\begin{aligned}
V_\mu \rightarrow V^\prime = UV_\mu U^\dagger + U \partial_\mu U^\dagger, \\
A_\mu \rightarrow A^\prime = U A_\mu U^\dagger \,.
\end{aligned}
\end{equation}
Meanwhile the nucleons transform as
\begin{equation}
    N \rightarrow N^\prime = U\,N,
\end{equation}
and
\begin{equation}
    D_\mu N \equiv \partial_\mu N + V_\mu N
        \rightarrow U(D_\mu N) \,.
\end{equation}

\subsection{Naive $\Lambda_{\chi SB}$ operator power counting}
\label{Naive power counting}

The \chiPTtwb  Lagrangian,
including all analytic quantum-loop effects 
for soft momenta ($\ll1$ GeV), 
can be written \cite{Manohar1984189, Georgi1993187}:
\begin{widetext}
\begin{eqnarray}
\label{L_chiPT_full}
  L_{\chi PT} =
     -\sum_{\substack{l,m,n\\l+m\geq1}}\!\!\!
     	 C_{lmn} f^2_\pi\Lambda^2_{\chi SB}
     \left ( \frac{\partial_\mu}{\Lambda_{\chi SB}}\right)^{\!\!m}
     \left ( \frac{\overline{N} N}{f_\pi^2\,\,\Lambda_{\chi SB}}\right)^{\!\!l} 
     \left ( \frac{m_{quark}}{\Lambda_{\chi SB}}\right)^{\!\!n} 
     f_{lmn}\left ( \frac{\pi_a}{f_\pi} \right),
\end{eqnarray}
\end{widetext}
where $f_{lmn}$ is an analytic function,
and the dimensionless constants $C_{lmn}$ 
are ${\mathcal O}(\Lambda_{\chi SB}^0)$ and, presumably,
$\sim 1$. 
As a power series in $\Lambda_{\chi SB}$
we take, self-consistently, $\Lambda_{\chi SB}\simeq 1$ GeV
 and, in higher orders,
reorder the non-relativistic perturbation expansion in $\partial_0$ 
to converge with large nucleon mass 
$m^N \approx \Lambda_{\chi SB}$ \cite{Weinberg1990288, Weinberg19913, Weinberg1992114}.

\subsection{The Chiral Symmetric Limit}

For the purposes of this paper, 
we retain from \eqref{L_chiPT_full} only terms of order 
$\Lambda_{\chi SB}$ and $\Lambda^0_{\chi SB}$, i.e., $1\leq m+l+n\leq2$.
We can further divide $L_{\chi PT}$
into a symmetric piece 
(i.e., with spontaneous 
breaking and massless Goldstones) 
and a symmetry-breaking piece 
(i.e., explicit breaking, arising from non-zero quark masses) 
generating three massive pNGB: 
\begin{equation}
	L_{\chi PT} = L^{Sym}_{\chi PT} + L^{Sym-Breaking}_{\chi PT} \,.
\end{equation}
In this paper, 
we are interested only in unbroken \chiPTtwb 
and so take $n=0$ in \eqref{L_chiPT_full}
\begin{eqnarray}
	L^{Sym-Breaking}_{\chi PT} = 0.
\end{eqnarray}
\noindent
We separate $L^{Sym}_{\chi PT}$ into 
pure meson terms, 
terms quadratic in baryons (i.e., nucleons), 
and four-baryon terms:
\begin{eqnarray}
	L^{Sym}_{\chi PT} &=& 
	L^{\pi;Sym}_{\chi PT} + L^{N;Sym}_{\chi PT} + L^{4-N;Sym}_{\chi PT} 
\end{eqnarray}
with (as in \eqref{Symmetric_Lagrangian}):
\be\ba
\label{Symmetric_Lagrangian_Appendix}
	L^{\pi; Sym}_{\chi PT} &= 
		\frac{f^2_\pi}{4}\Tr \partial_\mu \Sigma \partial^\mu \Sigma^\dagger, \\
	L^{N; Sym}_{\chi PT} &= 
		\overline{N} \left(i\gamma^\mu D_\mu -  \tildemone\right)N  \\
		&\quad \quad- g_A\overline{N}\gamma^\mu\gamma^5{A_\mu N },  \\
	L^{4-N;Sym}_{\chi PT} &\sim
		\frac{1}{f^2_\pi}\left(\overline{N}\gamma^{\mathscr A} N\right)
\left( \overline{N}\gamma_{\mathscr A} N\right) +++\,,
\ea\ee

As described below \eqref{Symmetric_Lagrangian},
the parentheses in the four-nucleon Lagrangian 
indicate the order of SU(2) index contraction,
and  $+++$  indicates that one should include 
all possible combinations of such contractions.
As usual,
  $\gamma^{\mathscr A} \equiv
  \left(1, \gamma^\mu,  i\sigma^{\mu\nu},  
  i\gamma^\mu \gamma^5,  \gamma^5\right)$,
for ${\mathscr A}=1,...,16$
(with 
$\sigma^{\mu\nu} \equiv \frac{1}{2}[\gamma^\mu,  \gamma^\nu ]$).
These are commonly referred to as 
scalar (S), vector (V), tensor (T), axial-vector (A), and pseudo-scalar (P)
respectively. 

In this paper, we will focus on the {\it Semi-Classical Symmetries} of chiral (i.e., spontaneously broken) SU(2) $\chi$PT. Nucleons are treated as quantized fermions.  Pions are classical fields: i.e., $\xi$, $V_\mu$, $A_\mu$, $U$, $\Sigma$, $\pi_a$, $R$, $L$ defined
in Subsection (\ref{NonLinearTransformationProperties})
are not quantized: their non-trivial commutation properties are entirely due to strong isospin.

\subsection{SU$(2)_L\times$SU$(2)_R$ invariant 4-nucleon contact interactions}
We focus on the 4-fermion terms in \eqref{Symmetric_Lagrangian_Appendix}.  We use the completeness relation for $2\times 2$ matrices:
\be
\ba
\delta_{cf}\delta_{ed} = 	{2}\sum_{B=0}^3 t^B_{cd} t^B_{ef}\,; \\
	\left[ {\vec t} \, ,\, U\left({\vec \pi}(x), r, l \right) \right] \neq 0;
\ea
\ee
with $t^B=(\tfrac{1}{2} I, {\vec t})$.  (We use Greek letters for relativistic spinor indices, and Roman letters for isospin indices.)
Both iso-scalar and iso-vector 4-nucleon contact interactions 
appear in the \SUtwLR invariant Lagrangian:
\begin{nalign}
\label{L4NSymmetricchiPT}
&L^{4-N;Sym}_{\chi PT} =\frac{C^{T=0}_{\mathscr A}}{f^2_\pi}( \overline{N}_a^\alpha\gamma^{{\mathscr A}\alpha\beta} N_a^\beta)( \overline{N}_b^\lambda \, \gamma_{\mathscr A}^{\lambda\sigma} N_b^\sigma) \\
&\quad\quad+\frac{C^{T=1}_{\mathscr A}}{f^2_\pi}( \overline{N}_a^\alpha\gamma^{{\mathscr A}\alpha\beta} N_b^\beta)( \overline{N}_b^\lambda\,\gamma_{\mathscr A}^{\lambda\sigma} N_a^\sigma)  \\
&=\frac{C^{T=0}_{\mathscr A}}{f^2_\pi}( \overline{N}_c^\alpha\,U^\dagger_{ca}\gamma^{{\mathscr A}\alpha\beta} U_{ad}N_d^\beta)( \overline{N}_e^\lambda\,U^\dagger_{eb}\gamma_{\mathscr A}^{\lambda\sigma} U_{bf}N_f^\sigma) \\
&\quad+
\frac{C^{T=1}_{\mathscr A}}{f^2_\pi}( \overline{N}_c^\alpha\,U^\dagger_{ca}\gamma^{{\mathscr A}\alpha\beta} U_{bd}N_d^\beta)( \overline{N}_e^\lambda\,U^\dagger_{eb}\gamma_{\mathscr A}^{\lambda\sigma} U_{af}N_f^\sigma) \\
&=\frac{C^{T=0}_{\mathscr A}}{f^2_\pi}( \overline{N}_c^\alpha\,\gamma^{{\mathscr A}\alpha\beta} N_c^\beta)( \overline{N}_e^\lambda\,\gamma_{\mathscr A}^{\lambda\sigma} N_e^\sigma) \\
&\quad+\frac{C^{T=1}_{\mathscr A}}{f^2_\pi}( \overline{N}_c^\alpha\,\gamma^{{\mathscr A}\alpha\beta} N_d^\beta)( \overline{N}_e^\lambda\,\gamma_{\mathscr A}^{\lambda\sigma} N_f^\sigma)
\delta_{cf}\delta_{ed} \\
&=\frac{C^{T=0}_{\mathscr A}}{f^2_\pi}( \overline{N}_c^\alpha\,\gamma^{{\mathscr A}\alpha\beta} N_c^\beta)( \overline{N}_e^\lambda\,\gamma_{\mathscr A}^{\lambda\sigma} N_e^\sigma) \\
&\quad+2\sum _{B=0}^3  \frac{C^{T=1}_{\mathscr A}}{f^2_\pi}( \overline{N}_c^\alpha\,t^B_{cd}\,\gamma^{{\mathscr A}\alpha\beta} N_d^\beta)( \overline{N}_e^\lambda\,t^B_{ef}\,\gamma_{\mathscr A}^{\lambda\sigma} N_f^\sigma)\,.
\end{nalign}

\section{4-nucleon contact interactions in \SchiNL's}

\subsection{Boson-exchange-inspired 4-nucleon contact interactions}
We  wish to study the expectation value of 
$L^{4-N;Sym}_{\chi PT}$ in the ground state of the chiral nuclear liquid (which we continue to represent with $\bran \, \ketn$).  Using \eqref{L4NSymmetricchiPT} we find:
\be
\ba
-L_{S\chi NL}^{BE} &\equiv \bran -L^{4-N;Sym}_{\chi PT} \ketn = \\
 &\sum_{{\mathscr A}} \frac{C^{T=0}_{\mathscr A}}{2f_\pi^2} \bran \overline{N}_c^\alpha\,\gamma^{{\mathscr A}\alpha\beta} N_c^\beta\, \overline{N}_e^\lambda\,\gamma_{\mathscr A}^{\lambda\sigma} N_e^\sigma \ketn \\
+&\sum_{{\mathscr A}B}\frac{C^{T=1}_{\mathscr A}}{f_\pi^2} \bran \overline{N}_c^\alpha\,t^B_{cd}\,\gamma^{{\mathscr A}\alpha\beta} N_d^\beta\, \overline{N}_e^\lambda\,t^B_{ef}\,\gamma_{\mathscr A}^{\lambda\sigma} N_f^\sigma\ketn\,
\ea\ee
In what follows we ignore any and all excited states 
and consider the effective Lagrangian:
\be\ba
\label{Boson_Exchange_Appendix_c}
&-L_{S\chi NL}^{BE} =\frac{1}{2f_\pi^2}\sum_{{\mathscr A}} \\
&\quad \times \left\{ C^{T=0}_{\mathscr A} \bran \overline{N}_c^\alpha\,\gamma^{{\mathscr A}\alpha\beta} N_c^\beta
\ketn\bran
\overline{N}_e^\lambda\,\gamma_{\mathscr A}^{\lambda\sigma} \,N_e^\sigma \ketn \right. \\
&\quad \quad \left. + 2\, \sum_B C^{T=1}_{\mathscr A}\bran \overline{N}_c^\alpha\,t^B_{cd}\,\gamma^{{\mathscr A}\alpha\beta} N_d^\beta
\ketn \right.\\
&\quad\quad\quad\quad \quad \left. \times \bran \overline{N}_e^\lambda\,t^B_{ef}\,\gamma_{\mathscr A}^{\lambda\sigma} N_f^\sigma \ketn \right\}.
\ea\ee
A useful identity is:
\be\ba
\label{Useful_Theorem}
&  \tfrac{1}{4}\bran \overline{N}_c^\alpha\,\gamma^{{\mathscr A}\alpha\beta} N_c^\beta
\ketn\bran
\overline{N}_e^\lambda\,\gamma_{\mathscr A}^{\lambda\sigma} N_e^\sigma \ketn  \\
&+   \bran \overline{N}_c^\alpha\, t_{3;cd}\,\gamma^{{\mathscr A}\alpha\beta} N_d^\beta
\ketn 
\bran \overline{N}_e^\lambda\,t_{3;ef}\,\gamma_{\mathscr A}^{\lambda\sigma} N_f^\sigma\ketn \\
&= \shalf \bran \overline{p}_c^\alpha\,\gamma^{{\mathscr A}\alpha\beta} p_c^\beta
\ketn\bran
\overline{p}_e^\lambda\,\gamma_{\mathscr A}^{\lambda\sigma} p_e^\sigma \ketn  \\
&  \quad+\shalf \bran \overline{n}_c^\alpha\,\gamma^{{\mathscr A}\alpha\beta} n_c^\beta
\ketn\bran
\overline{n}_e^\lambda\,\gamma_{\mathscr A}^{\lambda\sigma} n_e^\sigma \ketn.
\ea\ee

\subsection{Contact interactions that mimic hadronic boson-exchange}
Taking expectation values inside the \SchiNLb as in \eqref{Boson_Exchange_Appendix_c}, we obtain:
\label{Boson_Exchange_Appendix_c2}
\be
-L_{S\chi NL}^{BE} \simeq \frac{1}{2 f^2_\pi} \left(L^{T=0}_S+L^{T=0}_V+L^{T=1}_S + L^{T=1}_V\right)
\ee
where
\be\ba
\label{Boson_Exchange_Appendix_c3}
L^{T=0}_S &= C^{T=0}_{S} \left< \overline{N}_c^\alpha\, N_c^\alpha\right>\left< \overline{N}_e^\lambda\, N_e^\lambda \right>, \\
L^{T=0}_V &= C^{T=0}_{V} \left< \overline{N}_c^\alpha\,\gamma^{0;\alpha\beta} N_c^\beta\right>\left< \overline{N}_e^\lambda\,\gamma_{0}^{\lambda\sigma} N_e^\sigma\right>,  \\
L^{T=1}_V &= 2\,C^{T=1}_{S}\left\{ \frac{1}{4}\left< \overline{N}_c^\alpha\, N_c^\alpha\right> \left< \overline{N}_e^\lambda\, N_e^\lambda\right> \right.\\
&\quad +\left.\left< \overline{N}_c^\alpha\, {t}_{3;cd} \,N_d^\alpha\right>
\left< \overline{N}_e^\lambda\,{t}_{3;ef}\, N_f^\lambda\right> \right\}, \\
L^{T=1}_S &=2\,C^{T=1}_{V}\left\{ \frac{1}{4}\left< \overline{N}_c^\alpha\,\gamma^{0;\alpha\beta}\,N_c^\beta\right> \left<\overline{N}_e^\lambda\,\gamma_{0}^{\lambda\sigma} \,N_e^\sigma\right> \right.\\
&\quad +\left.\left< \overline{N}_c^\alpha\, {t}_{3;cd}\gamma^{0;\alpha\beta} \,N_d^\beta\right>
 \left< \overline{N}_e^\lambda\,{t}_{3;ef}\,\gamma_{0}^{\lambda\sigma} N_f^\sigma\right> \right\}.
\ea\ee
The factorization in $L_{S\chi NL}^{BE}$, 
 and its name, are inspired by a simple picture 
of forces carried by heavy hadronic-boson exchange, which is commonly envisioned in Walecka-like, nuclear-Skyrme and density-functional models;
i.e., we have integrated out the auxiliary fields:
\begin{itemize}
\item  Lorentz-scalar isoscalar $\sigma$, with chiral coefficient $C^{T=0}_{S}$\;;
\item  Lorentz-vector isoscalar $\omega_\mu$ with chiral coefficient $C^{T=0}_{V}$\;;
\item  Lorentz-scalar isovector ${\vec \delta}$, with chiral coefficient $C^{T=1}_{S}$;
\item  Lorentz-vector isovector ${\vec \rho}_\mu$, with chiral coefficient $C^{T=1}_{V}$.
\end{itemize}

To order $\Lambda^0_{\chi SB}$, 
the only 4-nucleon contact terms allowed 
by local SU(2) $\chi$PT symmetry 
are exhibited in \eqref{Boson_Exchange_Appendix_c} 
and \eqref{Boson_Exchange_Appendix_c2}. 
Note that isospin operators $\vec{t} = \half\vec{\tau}$  have appeared.
However, quantum-loop naive power counting 
requires inclusion of nucleon Lorentz-spinor-interchange interactions,
in order to enforce anti-symmetrization of fermion wavefunctions. 
These are the same order as direct interactions; i.e., $O(\Lambda^0_{\chi SB})$.
The empirical nuclear models 
of Manakos and Mannel \cite{Manakos1988223, Manakos1989481} 
were specifically built to include such spinor-interchange terms. 

Explicit inclusion of spinor-interchange terms yields a great technical advantage for the liquid approximation: it allows us to treat \SchiNL s in Hartree-Fock approximation, i.e., including fermion wave function anti-symmetrization, rather than in less-accurate Hartree approximation. 

\subsection{Contact-interactions, 
including spinor-interchange terms enforcing effective anti-symmetrization of fermion wavefunctions in the Hartree-Fock approximation}
\label{Contact_Spinor_Interchange}
In this section, we write an effective \SchiNLb Lagrangian for the four-nucleon contact interactions in terms of the ten independent chiral coefficients: $C^T_K$ with $K \in \{S,V,T,A,P\}$ and $T\in\{0,1\}$. 

For pedagogical simplicity, 
we first focus on the ``boson-exchange-inspired" terms, 
with power-counting contact-interactions of order $(\Lambda^0_{\chi SB})$.
``Direct" terms depend only on 
${C^{T=0}_{S}}$, ${C^{T=0}_{V}}$,${C^{T=1}_{S}}$, 	and ${C^{T=1}_{V}}$, 
because isoscalar (${C^{T=0}_{T}}$, ${C^{T=0}_{A}}$, and ${C^{T=0}_{P}}$) 
and isovector (${C^{T=1}_{T}}$, ${C^{T=1}_{A}}$, ${C^{T=1}_{P}}$) 
vanish when evaluated in the liquid.
``Spinor-interchange" terms depend all 10 coefficients after Fierz rearrangement.
(Such terms do not appear in the \chiPTtwb analysis 
of the deuteron ground state,
because it only has 1 proton and 1 neutron.)
The combination of direct and spinor-interchange terms 
(which we refer to below as ``Total'')
depends on all 10 coefficients. 

Because of the inclusion of spinor exchange terms,
Hartree treatment of the \SchiNLb Lagrangian 
is equivalent to Hartree-Fock treatment of the liquid. 
When building the semi-classical liquid quantum state, 
this enforces the anti-symmetrization of the fermion wavefunctions.
A crucial observation is that the resultant liquid 
depends on only four independent chiral coefficients:  
${\calC^{2}_{S}}$, ${\calC^{2}_{V}}$, ${\overline {\calC^{2}_{S}}}$, 
and ${\overline{\calC^{2}_{V}}}$.
These provide sufficient free parameters to balance 
the scalar attractive force carried by 
$\calC^{2}_{S}$ and $\overline {\calC^{2}_{S}}$ 
against the vector repulsive force carried by 
$\calC^{2}_{V}$ and $\overline {\calC^{2}_{V}}$ 
when fitting to the experimentally observed structure of ground state nuclei (as reflected, e.g., in the different signs in definition of $\calC^{2}_{V}$ and $\calC^{2}_{S}$ in \eqref{Walecka_parameters_V} and \eqref{Walecka_parameters_S}).

Motivated by the empirical success 
of Non-topological Soliton models 
we conjecture that 
excited-nucleon-inspired contact-interaction terms are small, 
and that the simple picture of  scalar attraction 
balanced against vector repulsion persists when including them. 
Such analysis is beyond the scope of this paper.

\subsubsection{Lorentz Vector (V) and Axial-vector (A) forces}

Proceeding in a similar manner for the vector and axial vector terms we find:
\begin{widetext}
\be\ba
\label{Vector_Boson_Exchange}
-L_{S\chi NL}^{V,A} \equiv -\left<L^{4-N;V,A} \right>& 
=	\frac{1}{2f_\pi^2}\sum_{{\mathscr A}=V,A} \biggl\{ C^{T=0}_{\mathscr A} \left< \overline{N}_c^\alpha\,\gamma^{{\mathscr A}\alpha\beta} \,N_c^\beta
\right>\left<
\overline{N}_e^\lambda\,\gamma_{\mathscr A}^{\lambda\sigma} \,N_e^\sigma \right> \\
&\qquad \quad +\,2 \sum_B C^{T=1}_{\mathscr A}\left< \overline{N}_c^\alpha\,t^B_{cd}\,\gamma^{{\mathscr A}\alpha\beta} \,N_d^\beta
\right> \left< \overline{N}_e^\lambda\,t^B_{ef}\,\gamma_{\mathscr A}^{\lambda\sigma} \, N_f^\sigma\right> \biggr\}
\ea\ee
\end{widetext}
\noindent
which is
\begin{widetext}
\be\ba
\label{Rewrite_vector_axial_vector}
-L_{S \chi NL}^{V,A} &=\frac{1}{2f^2_\pi} \sum_{{\mathscr A}=V,A}\biggl\{2 C^{T=0}_{\mathscr A} \left< \overline{p}_c^\alpha\,\gamma^{{\mathscr A}\alpha\beta} \,p_c^\beta
\right> \left<
\overline{n}_e^\lambda\,\gamma_{\mathscr A}^{\lambda\sigma} \,n_e^\sigma \right> \\
&+  \left[ C^{T=0}_{\mathscr A} + C^{T=1}_{\mathscr A} \right] \left[ \left< \overline{p}_c^\alpha\,\gamma^{{\mathscr A}\alpha\beta} \,p_c^\beta
\right> \left<
\overline{p}_e^\lambda\,\gamma_{\mathscr A}^{\lambda\sigma} \,p_e^\sigma \right> \right.
\left. + \left< \overline{n}_c^\alpha\,\gamma^{{\mathscr A}\alpha\beta}\, n_c^\beta
\right>\left<
\overline{n}_e^\lambda\,\gamma_{\mathscr A}^{\lambda\sigma} \,n_e^\sigma \right> \right] \biggr\}.
\ea\ee
\end{widetext}
\paragraph{Direct terms:}
The properties of \SchiNL s enable this expression to be written as:
\be\ba
\label{Simplify_direct_vector_axial_vector1}
-&L_{S \chi NL;D}^{V,A} =\frac{1}{2f^2_\pi} C^{T=0}_{V} \left\{ 2 \left< {p^{\dagger}}p\right> \left< {n^{\dagger}}n \right>  \right\} \\
& + \frac{1}{2f^2_\pi} \left[ C^{T=0}_{V} + C^{T=1}_{V}\right] \left\{\left< {p^{\dagger}}p \right>^2
+\left< {n^{\dagger}}n \right>^2\right\},
\ea\ee
where $\left< {p^{\dagger}}p\right> $ and  $\left< {n^{\dagger}}n\right>$ represent 
$\left< {p_c^{\alpha\dagger}}p_c^\alpha \right>$ and $\left< {n_e^{\lambda \dagger}}n_e^\lambda \right>$, respectively.

\paragraph{Spinor-interchange terms:}
After interchanging the appropriate spinors, normal ordering creation and annihilation operators, and Fierz re-arrangement, spinor-interchange contributions depend on $C^{T=0}_{V}$, $C^{T=0}_{A}$, $C^{T=1}_{V}$, and $C^{T=1}_{A}$. 
\be\ba
\label{Simplify_direct_vector_axial_vector2}
-&L_{S\chi NL;Ex}^{V,A} = \frac{1}{2f^2_\pi} \\
&\times\left[ -\left(C^{T=0}_{V} + C^{T=1}_{V} \right) + \left(C^{T=0}_{A} + C^{T=1}_{A} \right)  \right] \\ 
&\times\left\{  \left< {p_L^{\dagger}}p_L \right>^{\!2}\!+ \left< {p_R^{\dagger}}p_R \right>^{\!2}\!+ \left< {n_L^\dagger}n_L \right>^{\!2}\!+\!\left< {n_R^{\dagger}}n_R \right>^{\!2}
 \right\},
\ea\ee
where we have expanded the spinors $p$ and $n$ into left-handed and right-handed components via $p=p_L+p_R$ and $n=n_L+n_R$.

\paragraph{Total direct and spinor-interchange terms:}
Combining the direct and exchange terms yields:
\begin{nalign}
\label{Total_vector_axial_vector}
&-L_{S \chi NL;Total}^{V,A} =\frac{1}{f^2_\pi} C^{T=0}_{V} \left\{\left< {p^{\dagger}}p\right> \left< {n^{\dagger}}n \right>  \right\}\\
&+ \frac{C^{0}_{V} + C^{1}_{V}}{f^2_\pi} \left\{\left< {p_L^{\dagger}}p_L \right>\!
 \left< {p_R^\dagger}p_R \right>+\left< {n_L^\dagger}n_L \right>\!\left< {n_R^{\dagger}}n_R \right>\right\} \\
&+ \frac 
{C^{0}_{A} + C^{1}_{A}}{2f^2_\pi} 
\sum_{h=L,R}\left\{\left< {p_h^{\dagger}}p_h \right>^{\!2}
 + \left< {n_h^\dagger}n_h \right>^{\!2}
 \right\}.
\end{nalign}

\noindent
The reader should note the cancellation of the term
\be
\label{Vector_cancellation}
\frac{(C^{T=0}_{V} + C^{T=1}_{V})}{2f_\pi^2} 
\!\!\sum_{h=L,R}\left\{\left< {p_h^{\dagger}}p_h \right>^{\!2}
\!\!+ \left< {n_h^\dagger}n_h \right>^{\!2}
 \right\},
\ee
showing that vector-boson exchange 
cannot carry forces between same-handed fermion protons, 
or between same-handed fermion neutrons.

Significant simplification follows because 
\SchiNL s are defined to have equal left-handed and right-handed densities; i.e.,
\begin{nalign}
 \left< p^\dagger_L p_L\right>&=\left< p^\dagger_R p_R\right>=\half\left< p^\dagger p\right> \\
 \left< n^\dagger_L n_L\right>&=\left< n^\dagger_R n_R\right>=\half\left< n^\dagger n\right>.
\end{nalign}
Using \eqref{Walecka_parameters_V} the contribution of \eqref{Total_vector_axial_vector} 
to the Lorentz-spinor-interchange Lagrangian can be written:
\begin{nalign}
\label{Vector_axial_vector_chiral_coefficients}
-L_{S \chi NL;Total}^{V,A} =\tfrac{1}{2} \calC^{2}_{V} \left< {N^{\dagger}}N \right>^2 - {\overline {\calC^{2}_{V}}} \left< {N^{\dagger}}t_3 N \right>^2
\end{nalign}
with 
\be\ba
\label{Vector_nuclear_coefficients}
C^{V}_{200}&=C^{T=0}_{V} \\
-{\overline {C^{V}_{200}}} &= \shalf \left[ -C^{T=0}_{V}+C^{T=0}_{A} + C^{T=1}_{V}+ C^{T=1}_{A}\right].
\ea\ee
The crucial observation is that 
\eqref{Vector_axial_vector_chiral_coefficients} and \eqref{Vector_nuclear_coefficients}
depend on just {\em{two}} independent chiral coefficients, 
$\calC^{2}_{S}$ and ${\overline {\calC^{2}_{V}}}$, (or equivalently $C^{V}_{200}$ and ${\overline {C^{V}_{200}}}$), 
instead of four, 
while still providing sufficient free parameters 
to fit the vector repulsive force 
(i.e., within Non-topological Soliton, Density Functional 
and Skyrme nuclear models)
up to naive power-counting order $(\Lambda^0_{\chi SB})$, 
to the experimentally observed structure of ground state nuclei.

\subsubsection{Lorentz Scalar (S), Tensor (T) and Pseudo-scalar (P) forces}
Proceeding in a similar manner we define:
\be\ba
\label{Scalar_boson_exchange}
L_{S\chi NL}^{STP} & \equiv \left<L^{4-N;STP}_{\chi PT} \right> \\
\ea\ee
with
\begin{nalign}
&-L_{S \chi NL}^{STP}
=\frac{1}{2f_\pi^2}\sum_{{\mathscr A}=S,T,P} \\
&\quad\biggl\{ C^{T=0}_{\mathscr A} \left< \overline{N}_c^\alpha\,\gamma^{{\mathscr A}\alpha\beta} \,N_c^\beta \right>\left<
\overline{N}_e^\lambda\,\gamma_{\mathscr A}^{\lambda\sigma} \,N_e^\sigma \right> \\
&+2\, \sum_B C^{T=1}_{\mathscr A}\left< \overline{N}_c^\alpha\,t^B_{cd}\gamma^{{\mathscr A}\alpha\beta} \,N_d^\beta
\right> \left< \overline{N}_e^\lambda\,t^B_{ef}\gamma_{\mathscr A}^{\lambda\sigma} \,N_f^\sigma \right> \biggr\}
\end{nalign}
\noindent
which is:
\be\ba
\label{Rewrite_scalar_tensor_pseudoscalar}
&-L_{S \chi NL}^{STP} =\frac{1}{2f^2_\pi} \\
&\times \sum_{{\mathscr A}=S,T,P} \biggl\{2 \, C^{T=0}_{\mathscr A} \left<\overline{p}_c^\alpha\,\gamma^{{\mathscr A}\alpha\beta} \, p_c^\beta \right>\left<
\overline{n}_e^\lambda\,\gamma_{\mathscr A}^{\lambda\sigma} \,n_e^\sigma \right>  \\
& + \left[ C^{T=0}_{\mathscr A} + C^{T=1}_{\mathscr A} \right] \left[  \left< \overline{p}_c^\alpha\,\gamma^{{\mathscr A}\alpha\beta} \,p_c^\beta
\right>\left<
\overline{p}_e^\lambda\,\gamma_{\mathscr A}^{\lambda\sigma} \,p_e^\sigma \right> \right. \\
&\qquad \qquad + \left< \overline{n}_c^\alpha\,\gamma^{{\mathscr A}\alpha\beta} \,n_c^\beta
\right>\left. \left<
\overline{n}_e^\lambda\,\gamma_{\mathscr A}^{\lambda\sigma} \,n_e^\sigma \right> \right] \biggr\}
\ea\ee
\paragraph{Direct terms:}
The properties of \SchiNL s give
\be\ba
\label{Simplify_direct_scalar_tensor_pseudoscalar1}
-L_{S \chi NL;D}^{STP} &=\frac{1}{2f^2_\pi}\,C^{T=0}_{S}  \left< {\overline N}N\right> \left< {\overline N}N \right>  \\
& +\frac{1}{2f^2_\pi} \,C^{T=1}_{S} \left(  \left< {\overline p}p \right>
 \left< {\overline p}p \right> +\left< {\overline n}n \right>
 \left< {\overline n}n \right> \right).
\ea\ee

\paragraph{Spinor-interchange terms:}
Spinor-interchange contributions depend on six chiral coefficients: isoscalars $C^{T=0}_{S}$, $C^{T=0}_{T}$, $C^{T=0}_{P}$ 
and isovectors $C^{T=1}_{S}$, $C^{T=1}_{T}$, $C^{T=1}_{P}$. 
\be\ba
\label{Simplify_direct_scalar_tensor_pseudoscalar_2}
-L_{\chi NL;Ex}^{STP} &= \frac{1}{4f^2_\pi} 
\left[\left(C^{T=0}_{S} + C^{T=1}_{S} \right)\right.\\
&+6 \left(C^{T=0}_{T}+C^{T=1}_{T} \right) 
+ \left.\left(C^{T=0}_{P} + C^{T=1}_{P} \right)  \right] \\
&\times \left\{ 
\left< {\overline p_L}p_R \right>^2
 + \left< {\overline p_R}p_L \right>^2
 +\left< {\overline n_L}n_R \right>^2 
 + \left< {\overline n_R}n_L \right>^2
\right\}
\ea\ee
\paragraph{Total direct and spinor-interchange terms:}
As above, since \SchiNL s have equal left-handed and right-handed scalar densities by definition, 
the total direct and spinor-interchange contribution is considerably simplified:
\be\ba
\label{Scalar_tensor_pseudoscalar_chiral_coefficients}
&-L_{S \chi NL;Total}^{STP} =-\tfrac{1}{2} \calC^{2}_{S} \left<{\overline N}N \right>^2 
- \overline {\calC^{2}_{S}} \left<\overline {N}t_3 N \right>^2,
\ea\ee
where in \eqref{Walecka_parameters_S} and \eqref{Exchange_parameters_S} we have:
\be\ba
\label{Scalar_nuclear_coefficients}
C^{S}_{200}&=C^{T=0}_{S}, \\
-\overline {C^{S}_{200}} &= \half \left[\half C^{T=0}_{S}+ \frac{5}{2} C^{T=1}_{S} \right.\\
&+3\;\left.\left( C^{T=0}_{T} + C^{T=1}_{T}\right) + \left( C^{T=0}_{P}+  C^{T=1}_{P} \right) \right].
\ea\ee
\noindent
Once again we find that 
\eqref{Scalar_tensor_pseudoscalar_chiral_coefficients} and 
\eqref{Scalar_nuclear_coefficients}
depend on just {\em{two}} independent chiral coefficients, 
$C^{S}_{200}$ and ${\overline {C^{S}_{200}}}$, instead of six, 
while still providing sufficient free parameters 
to fit the scalar attractive force 
(i.e., within Non-topological Soliton, Density Functional 
and Skyrme nuclear models) 
up to naive power-counting order $\Lambda^0_{\chi SB}$, 
to the experimentally observed structure of ground state nuclei.

\section{Nucleon bi-linears and semi-classical nuclear currents in \SchiNL}
\label{Nuclear_current_section}
The structure of \SchiNLb suppresses various nucleon bi-linears:

\begin{itemize}
\item Vectors' space-components: because it is a 3-vector, parity odd and stationary 
\begin{eqnarray}
\bran \overline{N}_c^\alpha\,{\vec \gamma}^{\alpha\beta} N_c^\beta \ketn \sim  \bran {\vec k}\ketn \simeq 0 \quad \quad
\end{eqnarray}

\item Tensors: because the local expectation value of nuclear spin $<{\vec s}>=\half <{\vec \sigma}>\simeq 0$ 

\begin{enumerate}

\item $\sigma^{0j}$:
\be\ba
&\bran \overline{N}_c^\alpha\,\sigma^{0j;\alpha\beta} \,N_c^\beta\ketn \\
& \quad \quad =\bran \overline{N}_L \,\sigma^{0j} N_R \ketn +\bran \overline{N}_R\,\sigma^{0j} N_L
\ketn  \\
& \quad \quad = 2\bran \overline{N}_L 
{ \begin{aligned}
    \left [
    \begin{tabular}{ccc} 
    $0$ & ${\vec s}_j$ \\ 
    ${\vec s}_j$ & $0$      \\
    \end{tabular} \right]
  \end{aligned}}
 N_R \ketn \\
 & \quad \quad +2\bran \overline{N}_R
 { \begin{aligned}
 \label{pGoctetrepn3}
    \left [
    \begin{tabular}{ccc} 
    $0$ & ${\vec s}_j$ \\ 
    ${\vec s}_j$ & $0$      \\
    \end{tabular} \right]
  \end{aligned}}
  N_L
\ketn  \\
& \quad \quad \simeq 0
\ea\ee
                
\item $\sigma^{ij}$:
\be\ba
&\bran \overline{N}_c^\alpha\,\sigma^{ij;\alpha\beta} N_c^\beta\ketn   \\
& \quad \quad =\bran \overline{N}_L\,\sigma^{ij} N_R \ketn +\bran \overline{N}_R\,\sigma^{ij} N_L
\ketn   \\
& \quad \quad = -2i\epsilon_{ijk}\bran \overline{N}_L\, {\vec s}_k N_R \ketn  \\
& \quad \quad \quad -2i\epsilon_{ijk}\bran \overline{N}_R\, {\vec s}_k N_L\ketn \\
& \quad \quad \simeq 0
\ea\ee

\end{enumerate}

\item Axial-vectors: because ${p_L},{p_R}$ are equally represented in \SchiNL, as are ${n_L},{n_R}$:
\begin{nalign}
&\bran \overline{N}_c^\alpha\,\gamma^{A;\alpha\beta} N_c^\beta
\ketn  \\
&\quad \quad =\bran \overline{N}_L\,\gamma^\mu \gamma^5 N_L \ketn +\bran \overline{N}_R\,\gamma^\mu \gamma^5 N_R
\ketn  \\
&\quad \quad = -\bran \overline{N}_L\,\gamma^\mu N_L\ketn +\bran \overline{N}_R\, \gamma^\mu N_R
\ketn  \\
&\quad \quad \simeq 0 
\end{nalign}

\item Pseudo-scalars: because  \SchiNLb are of even parity
\begin{nalign}
&\bran \overline{N}_c^\alpha\,\gamma^{P;\alpha\beta} N_c^\beta
\ketn  \\
& \quad \quad =\bran \overline{N}_R\,\gamma^5 N_L \ketn +\bran \overline{N}_L\,\gamma^5 N_R
\ketn  \\
&\quad \quad = -\bran \overline{N}_R \, N_L \ketn +\bran \overline{N}_L\, N_R
\ketn  \\
&\quad \quad \simeq 0 
\end{nalign}
\end{itemize}

Therefore, various Lorentz and isospin representations are suppressed in \SchiNL s. In summary: Isoscalars
\begin{nalign}
\bran \overline{N}_c^\alpha\, N_c^\alpha \ketn &\neq 0,\\
\bran \overline{N}_c^\alpha\,\gamma^{0;\alpha\beta} N_c^\beta \ketn &\neq 0, \\
\bran \overline{N}_c^\alpha\,{\vec \gamma}^{\alpha\beta} N_c^\beta \ketn &\simeq 0, \\
\bran \overline{N}_c^\alpha\,\gamma^{T;\alpha\beta} N_c^\beta \ketn &\simeq 0, \\
\bran \overline{N}_c^\alpha\,\gamma^{A;\alpha\beta} N_c^\beta \ketn &\simeq 0, \\
\bran \overline{N}_c^\alpha\,\gamma^{P;\alpha\beta} N_c^\beta \ketn &\simeq 0.
\end{nalign}
and Isovectors
\begin{nalign}
 \bran \overline{N}_c^\alpha\, {t}^\pm_{cd}\gamma^{{\mathscr A}\alpha\beta} N_d^\beta \ketn &= 0, \\
 \bran \overline{N}_c^\alpha\, {t}^3_{cd}N_d^\alpha \ketn &\neq 0, \\
 \bran \overline{N}_c^\alpha\, {t}^3_{cd}\gamma^{0;\alpha\beta} N_d^\beta \ketn &\neq 0, \\
 \bran \overline{N}_c^\alpha\, {t}^3_{cd}{\vec \gamma}^{\alpha\beta} N_d^\beta \ketn &\simeq 0, \\
 \bran \overline{N}_c^\alpha\, {t}^3_{cd}\gamma^{T\alpha\beta} N_d^\beta \ketn &\simeq 0, \\
 \bran \overline{N}_c^\alpha\, {t}^3_{cd}\gamma^{A\alpha\beta} N_d^\beta \ketn &\simeq 0, \\
 \bran \overline{N}_c^\alpha\, {t}^3_{cd}\gamma^{P\alpha\beta} N_d^\beta \ketn &\simeq 0.
\end{nalign}
Now form the semi-classical nuclear currents
\begin{nalign}
  \label{Nucleon_bilinears_2}
  J^\mu_k &= \overline{N}\,\gamma^\mu t_k N, \quad k = 1, 2, 3 \\
  J^\mu_\pm &= J^\mu_1 \pm iJ^\mu_2 = \left \{ \begin{tabular}{c} $\overline{p}\gamma^\mu n$ \\ $\overline{n}\gamma^\mu p$ \end{tabular} \right\}, \\
  J^\mu_3 &= \frac{1}{2}\left ( \overline{p}\gamma^\mu p - \overline{n}\gamma^\mu n \right), \\
  J^\mu_8 &= \frac{\sqrt{3}}{2}\left ( \overline{p}\gamma^\mu p + \overline{n}\gamma^\mu n \right), \\
   J^\mu_{QED} &= \frac{1}{\sqrt{3}}  J^\mu_8 + J^\mu_3 = \overline{p}\gamma^\mu p \\
     J^\mu_{Baryon} &= \frac{2}{\sqrt{3}}  J^\mu_8  = \overline{p}\gamma^\mu p +  \overline{n}\gamma^\mu n, \\
  J^{5\mu}_k &= \overline{N}\gamma^\mu\gamma^5 t_k N, \quad
  k = 1, 2, 3 \\
  J^{5\mu}_\pm &= J^{5\mu}_1 \pm iJ^{5\mu}_2 = \left \{ \begin{tabular}{c} $\overline{p}\gamma^\mu \gamma^5 n$ \\ $\overline{n}\gamma^\mu  \gamma^5 p$ \end{tabular} \right\},  \\
  J^{5\mu}_{3} &= \frac{1}{2}\left ( \overline{p}\gamma^\mu \gamma^5 p - \overline{n}\gamma^\mu \gamma^5 n \right), \\
  J^{5\mu}_{8} &= \frac{\sqrt{3}}{2}\left ( \overline{p}\gamma^\mu \gamma^5 p + \overline{n}\gamma^\mu \gamma^5 n \right).
\end{nalign}

\SUtwLR nuclear currents within \SchiNLb are obedient to its semi-classical symmetries.  Thus we have:
\begin{equation}
\begin{aligned}
\label{Vanishing_currents_appendix}
\bran {J}^{\mu}_\pm \ketn &= \bran {J}^{\mu, 5}_\pm \ketn =\bran \partial_\mu{J}_{\pm}^\mu \ketn = 
\bran \partial_\mu{J}^{\mu, 5}_\pm \ketn=0,
\ea\ee
and
\begin{nalign}
\label{Approx_vanishing_currents_appendix}
\bran \partial_\mu{J}_{3}^\mu \ketn, \bran {J}^{\mu,5}_3 \ketn, \bran {J}^{\mu,5}_8 \ketn &\simeq 0, \\
\bran {J}^{\mu = 1,2,3}_3 \ketn, \bran {J}^{\mu = 1,2,3}_8 \ketn &\simeq 0, \\
\bran \partial_\mu{J}_{8}^\mu \ketn, \bran \partial_\mu{J}_{Baryon}^\mu \ketn, \bran \partial_\mu{J}_{QED}^\mu \ketn &\simeq 0, \\
\bran {J}^{\mu = 1,2,3}_{Baryon} \ketn, \bran {J}^{\mu = 1,2,3}_{QED} \ketn &\simeq 0,\\
\end{nalign}
\begin{nalign}
\frac{1}{\sqrt{3}}\bran \partial_\mu{J}^{\mu, 5}_8 \ketn &\propto \bran \bigl(m^N+ \widehat{{C_{200}^S}} \bigr) {\gamma}^{5}  \ketn \sim \eta \simeq 0,  \\
\tfrac{1}{2}\bran  \partial_\mu{J}^{\mu, 5}_3 \ketn &\propto \bran \bigl(m^N+\widehat{{C_{200}^S}} \bigr) {\gamma}^{5} t_3 \ketn \sim \pi_3 \simeq 0.  \\
\end{nalign}
The remaining non-zero contributions to the currents are:
\be\ba
\bran {J}^{0}_{Baryon} \ketn &\neq 0 ;\\
\bran {J}^{0}_3 \ketn &\neq 0;\\
\bran {J}^{0}_8 \ketn &\neq 0;  \\
\bran {J}^{0}_{QED} \ketn &\neq 0;  \\
\end{aligned}
\end{equation}
\section{Thomas-Fermi non-topological solitons and the semi-empirical mass formula}
\label{Thomas_Fermi}
We are interested here in semi-classical solutions to \eqref{Dirac_equation},
identifiable as quantum chiral nucleon liquids, 
that are, for reasons laid out in the main body of the paper:
	in the ground state, spin zero,
	spherically symmetric,  
	and even-even
	(i.e., have an even number of protons and of neutrons).
We employ  relativistic mean-field point-coupling 
	Hartree-Fock and Thomas-Fermi approximations,
ignoring the anti-nucleon sea.

We seek solutions that are static, homogeneous and isotropic. 
Given the absence of any surface terms at the order $\Lambda_{\chi SB}^0$ in chiral symmetry breaking to which we are working,
we avoid the {\it ad hoc} imposition of such terms.
We therefore impose the condition that the pressure vanishes everywhere,
rather than just at the surface of a finite ``liquid drop.''
Our finite Static$\chi NL$  nuclei therefore resemble "ice cream balls scooped from an infinite vat \cite{BogutaPrivate1993}, more than they do conventional liquid drops (which have surface tension). 

The Thomas-Fermi  approximation
replaces the neutrons and protons 
with homogeneous and isotropic expectation values 
over free neutron and proton spinors,
with (for $j=n$ and $p$)
effective reduced mass $m^j_*$,
3-momentum $\vec{k}^j$, 
energy $E^j = \sqrt{(\vec{k}^j)^2+(m^j_*)^2}$, 
and zero spin.
Most of these vanish because of the 
absence of any preferred direction for spin or momenta in Static$\chi NL$:
\be\ba
\label{liquid_means}
	\overline{n} n 
		&\rightarrow \langle \overline{n}n \rangle 
	  = \frac{m^n_*}{E_n}, \\
	\overline{n}\left(\gamma^0,\vec\gamma\right) n 
		&\rightarrow \langle\overline{n}(\gamma^0,\vec\gamma)n\rangle 
	  = (1,\vec0 ), \\
	\overline{n}\left(\sigma^{0j},\sigma^{ij}\right)n 
		&\rightarrow \langle\overline{n}
			\left(\sigma^{0j},\sigma^{ij}\right)n\rangle
	  = 0,\\
	\overline{n}\left(\gamma^0,\vec\gamma\right)\gamma^5n 
		&\rightarrow 
	    \langle\overline{n}\left(\gamma^0,\vec\gamma\right)\gamma^5n\rangle 
	  = 0, \\
	\overline{n}\gamma^5n &\rightarrow \langle \overline{n}\gamma^5 n \rangle = 0,
\ea\ee
and similarly for the proton. 
To simplify our notation, 
we drop the $\langle\cdots\rangle$ 
in the remainder of this appendix. 

Within the liquid drop, 
 the baryon number density 
\begin{equation}
  N^\dagger N = p^\dagger p + n^\dagger n\,,
\end{equation} 
and scalar density 
\begin{equation}
  \overline{N}N = \overline{p}p + \overline{n}n\,.
\end{equation} 
The neutron contributions to these densities are:
\begin{nalign}
n^\dagger n &= 2 \int_{0}^{k^n_F} \frac{d^3k}{(2\pi)^3} 
= \frac{(k^n_F)^3}{3\pi^2}, \\
\overline{n}n &= 2 \int_{0}^{k^n_F} \frac{d^3k}{(2\pi)^3}
                  \frac{m^n_*}{\sqrt{k^2+(m^n_*)^2}}, \\
    &= \frac{m^n_*}{2\pi^2}\left (k^n_F\mu^n_* 
             - \frac{1}{2}(m^n_*)^2 \ln \left (
                \frac{\mu^n_*+k^n_F}{\mu^n_*-k^n_F} \right) \right),
\label{Neutron_densities}
\end{nalign}
with
\be\ba
  m^n_* &\equiv \tildemn 
    		+{\calC^2_{S}}\overline{N}\,N -\tfrac{1}{2} \overline{\calC^S_{200}} (\overline{n}n-\overline{p}p), \\
  \mu^n_* &\equiv \sqrt{(k^n_F)^2 + (m^n_*)^2}.
\label{Reduced_mass_equation}
\ea\ee
The equivalent proton contributions are obtained 
by straightforward substitution of $n \leftrightarrow p$.

It is convenient to define:
\be\ba
  \epsilon^{\int n}
  	&\equiv2 \int_{0}^{k^n_F} \frac{d^3k}{(2\pi)^3} 
\sqrt{k^2 + (m^n_*)^2}, \\
&= \frac{3}{4}\mu^n_*n^\dagger n + \frac{1}{4}m^n_*\overline{n}n,
\ea\ee
\be\ba
  P^{\int n} 
&\equiv   2 \int_{0}^{k^n_F} \frac{d^3k}{(2\pi)^3} 
\frac{k^2}{3\sqrt{k^2 + (m^n_*)^2}}, \\
&= \frac{1}{4}\mu^n_*n^\dagger n-\frac{1}{4}m^n_*\overline{n}n,
\ea\ee
and equivalently for protons. 
These look conveniently like the neutron and proton
energy density and pressure, and indeed:
\begin{nalign}
\epsilon^{\int n} - 3 P^{\int n} &= m^n_* {\bar n} n,  \\
\epsilon^{\int n} + P^{\int n} &= \mu^n_* n^\dagger n	.
\end{nalign}
The actual nucleon energy density and pressure 
are properly constructed from the stress-energy tensor: 
\begin{eqnarray}
  \left ( T^{N}_{\chi PT} \right)^{\mu\nu} &\!\!=& 
    \frac{\partial L^{N}_{\chi PT}}
    {\partial \left ( \partial_\mu N \right)}\partial^\nu N 
    - g^{\mu\nu}L^{N}_{\chi PT},
\end{eqnarray}
with
\be\ba
\epsilon^N &\equiv \left ( T^{N}_{\chi PT}\right)^{00}, \\
  P^N &\equiv \frac{1}{3}\left ( T^{N}_{\chi PT}\right)^{jj}.
\ea\ee
The total nucleon energy and pressure are thus:
\noindent
\begin{widetext}
\be
\ba
\epsilon^N &= \epsilon^{\int p} + \epsilon^{\int n} 
    +\frac{1}{2} \left (\calC_V^2 (N^\dagger N)^2 
    - \frac{\overline{\calC^2_{V}}}{2}
        (p^\dagger p - n^\dagger n)^2 \right)
  +\frac{1}{2}\left (\calC^2_S (\overline{N}N)^2 
     + \frac{\overline{\calC^S_{200}}}{2}
        \left (\overline{p}p -\overline{n}n \right)^2 \right),\\
P^N &= P^{\int p} + P^{\int n} + 
\frac{1}{2} \left (\calC^2_V (N^\dagger N)^2
-\frac{\overline{\calC^2_{V}}}{2}
(p^\dagger p - n^\dagger n )^2 \right)
-\frac{1}{2}\left (\calC^2_S (\overline{N}N)^2 + \frac{\overline{\calC^2_{S}}}{2}(\overline{p}p - \overline{n}n)^2 \right).\\
\ea
\ee
\end{widetext}
Using 
\be\ba
  \mu^{n}_B &\equiv \mu^n_* 
    + \calC^2_V N^\dagger N - \overline{\calC^2_V} (n^\dagger n - p^\dagger p),  \\
  \mu^{p}_B &\equiv \mu^p_* 
    + \calC^2_V  N^\dagger N + \overline{\calC^2_V} (n^\dagger n - p^\dagger p),  
\ea\ee
\noindent
it follows that $\epsilon^N$ and $P^N$ are related by the baryon number densities: 
\begin{equation}
\epsilon^N + P^N = \mu^p_B \,\, p^\dagger p + \mu^n_B \,\, n^\dagger n \,.
\end{equation}
\noindent
The objects of our calculations are therefore the six quantities:
$\mu^{n,p}_{B}$, $m^{n,p}_{*}$, and $k^{n,p}_{F}$.
These are, respectively, 
the chemical potential, 
reduced mass,
and Fermi-momentum for neutrons and protons.

\subsection{$Z=N$ heavy nuclei in the chiral symmetric limit}
\label{ZEqualNappendix}
To calculate binding energies, we work in the chiral symmetric limit, $m_p=m_n$: e.g. zero electromagnetic breaking,
and $m_8 = \half (m_p+m_n)$.
We first study the case $Z=N$,
so $m^n_*=m^p_*\equiv m_*$ for equal numbers of protons and neutrons.
We search for 
a solution of the chiral-symmetric liquid equations that has $P^N=0$.
In this simple case, 
$\mu_B^{p}=\mu_B^{n}\equiv\mu_B$,
$\mu_*^{p}=\mu_*^{n}\equiv\mu_*$,
$m_{p}=m_{n}\equiv m_N$,
and
$k_{Fn}=k_{Fp}\equiv k_F$.
Thus 
\begin{equation}
	\label{kFfornum}
	k_F = \sqrt{\mu_*^2-m_*^2}\,.
\end{equation}
\noindent
We also have
$n^\dagger n=p^\dagger p=\half  N^\dagger N$, and
$\overline{n}n=\overline{p}p=\half \overline{N}N$.
We are therefore able to write the baryon density as:
\begin{equation}
\label{mu_minus_mustar_for_num}
N^\dagger N = \frac{\mu_B - \mu_*}{\calC^2_V},
\end{equation}
and the scalar density as: 
\begin{equation}
\label{m_minus_mstar_for_scalar}
\overline{N}N = \frac{m_N-m_*}{\calC^2_S},
\end{equation}
where, to make connection to Walecka's model of nuclear matter,
we use ${\calC^2_V}$ and ${\calC^2_S}$ defined in \eqref{Walecka_parameters_S} and \eqref{Walecka_parameters_S}, respectively.
\noindent
The baryon number and scalar densities are simply twice the values in \eqref{Neutron_densities}; i.e,: 
\begin{nalign}
\label{Nucleon_densities}
N^\dagger N \,&=\,\frac{2 k_F^3}{3\pi^2}, \\
\overline{N} N \,&=\,
\frac{ m_*}{\pi^2} 
\left ( \mu_* k_F -m^2_* \ln \left [ \frac{\mu_* + k_F}{m_*} \right] \right).
\end{nalign} 
\noindent
The fermion pressure is now:
\be\ba
\label{PPsifornum}
P^{N} &=
\frac{1}{4}\left[ \mu_B N^\dagger N +\calC_V^2 (N^\dagger N)^2 \right. \\
&\qquad\qquad\left. -m_N \overline{N}N-\calC_S^2 ({\overline N } N)^2 \right].
\ea\ee
To these six equations \eqref{kFfornum}-\eqref{PPsifornum}
in the seven variables 
$k_F$, $\mu_*$, $\delta_\mu\equiv\mu-\mu_*$, $m_*$,
$\overline{N} N$, $N^\dagger N$ 
and $P^N$, 
we add the physical condition that the Static$\chi NL$ non-topological soliton pressure vanish internally, in order that it remain stable when immersed in the physical vacuum:
\begin{equation}
P^N=0\,,
\end{equation}
eliminating $P^N$ as a free variable.
\begin{figure}[htp]
\centering
\centering
\includegraphics[width=3 in,scale=.5]{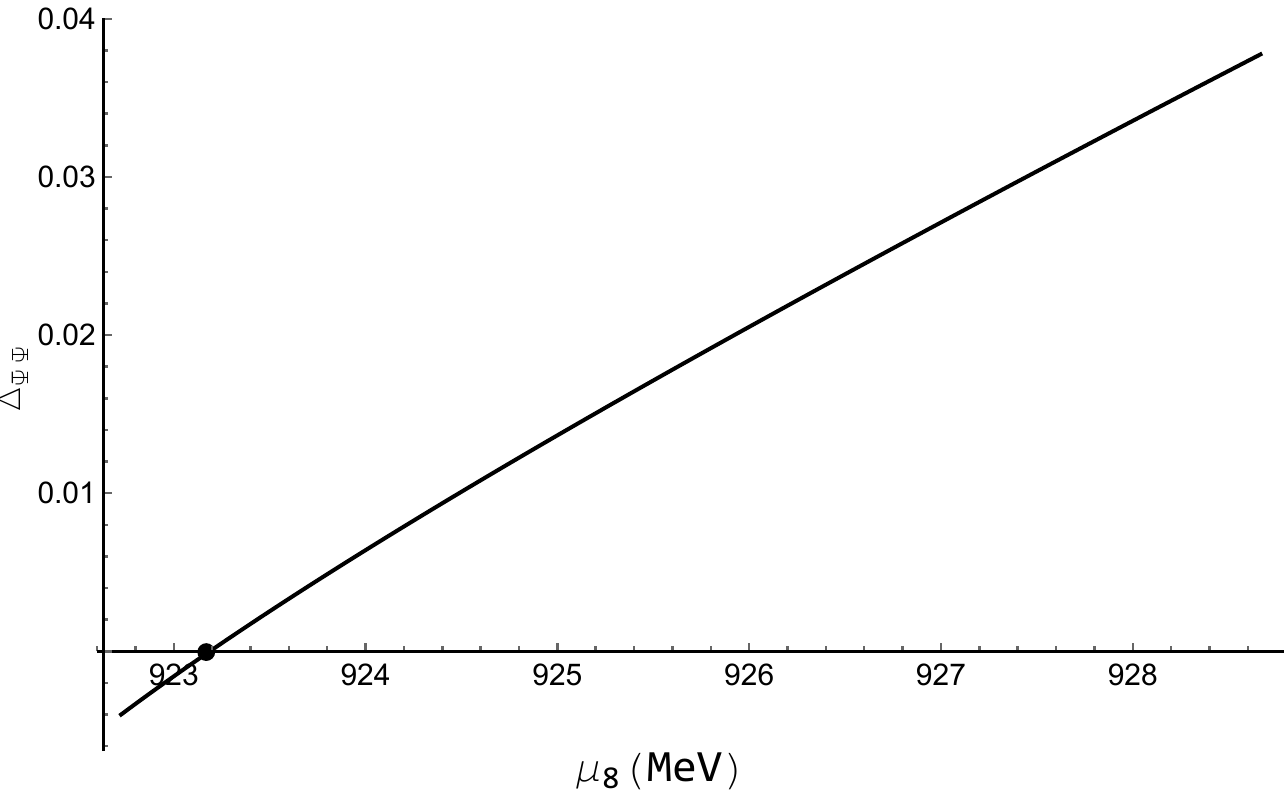}
\caption{$\Delta_{{\bar N} N}$ (cf. \eqref{root_condition}) as a function of baryon chemical potential $\mu_B$ for $\calC^2_V = 222.65$ GeV$^{-2}$ 
and $\calC^2_S = 303.45$ GeV$^{-2}$,
the Chin and Walecka values \cite{Chin197424} equivalent to ours.  A solution of the complete set of $Z=N$ chiral-symmetric pressure-less liquid equations 
must have $\Delta_{{\bar N}\,N}=0$, and thus is found at $\mu_B\simeq 923.17$ MeV, where the curve intersects the $\mu_B$ axis.  This value equals the Chin-Walecka value shown as a black dot.}
\label{fig:muB}
\end{figure}
Equations \eqref{kFfornum}-\eqref{Nucleon_densities}
can be solved analytically to give 
$k_F$,$\mu_*$, $N^\dagger N$ and $\overline N N$ 
as functions of $m_*$ and  $\delta_\mu$:
\be\ba
\label{kFmustarofmstar}
k_F &= \left(\frac{3\pi^2}{2}\frac{\delta_\mu}{\calC_V^2} \right)^{1/3}  \\
\mu_* &= \sqrt{k_F^2 + m_*^2}.
\ea\ee
Equation \eqref{PPsifornum}, with $P^N=0$, then becomes
a quartic equation for $m_*$ in terms of $\delta_\mu$:
\begin{equation}
\begin{aligned}
0 =& m_*^2  + \left( \frac{3\pi^2\delta_\mu}{2\calC_V^2}  \right)^{2/3}  \\
&- \left[
\frac{\calC_V^2}{\calC_S^2}\frac{(m_N-m_*)(2m_N-m_*)}{\delta_\mu} -2\delta_\mu
\right]^2,
\end{aligned} 
\end{equation}
which has up to four roots
$m_*(\delta_\mu;\calC_S^2,\calC_V^2)$,
for every value of $\delta_\mu$, $\calC_S^2$, and $\calC_V^2$\footnote{But only one of these four roots might be an infinite Static$\chi NL$, and then only if it were the $P^N\rightarrow0$ limit of a {\it finite} Walecka non-topological soliton. Those solitons satisfy {\it "Newtonian roll-around-ology"} \cite{Lynn1993281,Bahcall199067,Bahcall1989606,Selipsky1989430,Lynn1989465,Bahcall1998959,Bahcall1992} where the mean field nucleons move within a dynamic $\sigma$ field. $P^N_{Internal}\neq0$ and $P^N_{External}=0$ are then connected by the dynamic $\sigma$ surface.
}.
To be an actual solution of the complete set of 
$Z=N$ chiral-symmetric pressure-less liquid equations,
the root must also satisfy  \eqref{m_minus_mstar_for_scalar} and the second of 
\eqref{Nucleon_densities}; i.e.,
\begin{equation}
\label{root_condition}
\Delta_{{\bar N} N}  \equiv 
1 - \frac{ \calC_S^2 }{m_N-m_*}\,\overline{N} N
= 0
\end{equation}
where we use \eqref{kFmustarofmstar} for 
$k_F(\delta_\mu;\calC_V^2)$ 
and $\mu_*(\delta_\mu;\calC_S^2,\calC_V^2)$).

Now $m_N/f_\pi\approx939/93\approx10.10$, 
but in principle $\calC_S^2$ and $\calC_V^2$
are free parameters.
For given values of  $\calC_S^2$ and $\calC_V^2$,
we must search for a value of $\delta\mu$ 
such that \eqref{root_condition} holds.
The existence of such a value of $\delta\mu$ 
is not assured for arbitrary values of $\calC_S^2$ and $\calC_V^2$.

\begin{figure}[htp]
\includegraphics[trim={0 0 0 0},scale=.85]{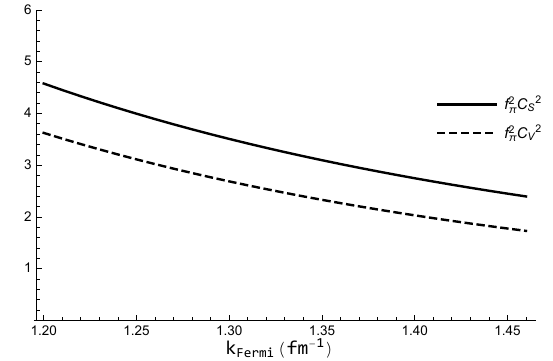}
\centering
\caption{Plot of $f_{\pi}^2 C_S^2$ and $f_{\pi}^2 C_V^2$ against the Fermi level in inverse $F$.  The calibration used a bulk binding energy $E_{Vol}=15.75$ MeV.}
\centering
\label{CS2_CV2_Versus_kF}
\end{figure}

Fitting to experimental values, Chin and Walecka found that their parameters 
$\calC^2_V = 222.65$ GeV$^{-2}$ and $\calC^2_S = 303.45$ GeV$^{-2}$.   
In Figure \ref{fig:muB}, we show that there does indeed exist a pressure-less chiral-symmetric nuclear liquid for  
$\calC_S^2$  and $\calC_V^2$ equal to the Chin and Walecka \cite{Chin197424} values. 
Furthermore, the inferred value of the baryon chemical
potential is $923.17$ MeV, and is consistent with Chen and Walecka's value.
Figure \ref{CS2_CV2_Versus_kF} shows representative values of $\calC_V^2$ and $\calC_S^2$ for different values of $k_F$ using the first approach.
Remarkably, 
we can now understand Chin and Walecka's nuclear matter to {\em be} a pressure-less chiral-symmetric nuclear liquid.
We also perhaps thereby gain some insight into the relative insensitivity of nuclear properties to pion properties.

\subsection{$Z\neq N$ heavy nuclei in the chiral-symmetric limit}
\label{ZneqNappendix}
Here we outline the analytic and numerical treatment of the case where $Z \neq N$ in the chiral limit.  
The approach may be summarized as follows:
\begin{enumerate}
\item The starting point is the zeroth order solution for the case $Z=N$ which determines the coupling constants $C_S^2$ and $C_V^2$ for a given Fermi level and binding energy as in the previous section.
\item All proton and neutron specific quantities are expanded in a Taylor series;
\item The general rule is: quantities vanishing in zeroth order have a first order variation while those not vanishing in zeroth order have only a second order variation; thus all terms up to second order must be retained;
\item The vanishing of pressure to second order provides an additional equation which allows all variations to be expressed in terms of the first order change in density only;
\item Since there appears no way to infer separately the value of $\overline{C_{S}^2}$ we follow Niksic and co-workers \cite{Niksic2008.034318} 
and set this constant to zero.  This leads to significant simplification.  In particular, changes in the proton and neutron reduced masses are equal in first order.    
\item We then solve for $\overline{\calC_{V}^2}$ by setting the asymmetry energy of the liquid model to the second order variation in the Thomas-Fermi energy.
\end{enumerate}
\noindent
In this section we use the following notation for the number and scalar densities:
\begin{equation}
\begin{aligned}
\rho_p \equiv & p^{\dagger} p;  \quad  \rho_n \equiv & n^{\dagger} n;  \quad  \rho_{\pm} =& \, \rho_p \pm \rho_{n} \,. \\
\rho_{Sp} \equiv & \overline{p} \, p;  \quad  \rho_{Sn} \equiv & \overline{n} \,n; \quad  \rho_{S\pm} =& \,\rho_{Sp} \pm \rho_{Sn}\,.
\label{Density_notation}
\end{aligned}
\end{equation}
We define the changes in densities as follows:
\begin{equation}
\begin{aligned}
d\rho_p - d\rho_{n} = \epsilon d \rho_{-}  \\
d\rho_p + d\rho_{n} = \epsilon^2 d \rho_{+}
\label{Number_density_expansion}
\end{aligned}
\end{equation}
where $\epsilon$ is merely a placeholder for the order of the variation.  It then follows that:
\begin{equation}
\begin{aligned}
\rho_p = & \tfrac{1}{2}\rho_+ + \frac{\epsilon}{2} d \rho_{-} + \frac{\epsilon^2}{2} d \rho_{+}, \\
\rho_n = & \tfrac{1}{2}\rho_+ - \frac{\epsilon}{2} d \rho_{-} + \frac{\epsilon^2}{2} d \rho_{+}.
\label{PN_density_expansion}
\end{aligned}
\end{equation}
Since the number density for each species is given by the first of \eqref{Neutron_densities} we get the following expansions for the Fermi levels:
\begin{equation}
\begin{aligned}
\delta k_{Fp} - \delta k_{Fn} = & \epsilon \frac{2 k_{F}}{3 \rho_+} \delta \rho_{-}, \\
\delta k_{Fp} + \delta k_{Fn} = & \epsilon^2 \left( \frac{2 k_{Fp}}{3 \rho_+}\delta \rho_{+} -\frac{2 k_F}{9 \rho_+^2} \delta \rho_{-}^2 \right).
\end{aligned}
\end{equation}
\noindent
It follows that:
\begin{equation}
\begin{aligned}
m_{*8} \,\equiv& \,\tfrac{1}{2}(m_{*p}+m_{*n}) = m_N - {C}_{S}^2 \, \rho_{S+}  \\
m_{*3} \,\equiv& \,\tfrac{1}{2}(m_{*p}-m_{*n}) = \frac{\overline{\calC_{S}^2}}{2} \, \rho_{S-},
\label{Scalar_density_equations}
\end{aligned}
\end{equation}
where we used the second of \eqref{Walecka_parameters_S}.  We also define:
\begin{equation}
\begin{aligned}
\mu_{*8} =& \tfrac{1}{2}(\mu_{*p}+\mu_{*n}),  \\
\mu_{*3} =& \tfrac{1}{2}(\mu_{*p}-\mu_{*n}),
\end{aligned}
\end{equation}
with $\mu_{*n,p}$ as in \eqref{Reduced_mass_equation}.
We now enforce $\overline{\calC_{S}^2}=0$: it follows immediately from the second of \eqref{Scalar_density_equations} that $m_{*3} = \delta m_{*3}=0$ with considerable simplification.  First, $\delta \mu_{*3}$ is a linear function of $\delta \rho_{-}$ only; i.e.,
\begin{eqnarray}
\delta \mu_{*3}=\frac{\pi^2}{2 \, k_{F} \, \mu_{*8}} \,\delta \rho_{-}.
\label{Chemical_potential_variations_3}
\end{eqnarray}
Second, $\delta \mu_{*8}$ is also simplified:
\begin{equation}
\begin{aligned}
\delta \mu_{*8}=&\frac{m_{*8}}{\mu_{*8}} \,\delta m_{*8} + \frac{\pi^2}{2 \mu_{*8} k_{Fp}} \delta \rho_{+}\\
&-\frac{\pi^4}{8 \,k_F^4 \,\mu_{*8}^3} \left(m_{*8}^2 + 2 k_F^2 \right) \delta \rho_{-}^2.
\label{Chemical_potential_variations_8}
\end{aligned}
\end{equation}
(As noted, $\delta \mu_{*3} $ is first order, while $\delta \mu_{*8}$ is second order.)  
The variation in the first of \eqref{Scalar_density_equations} gives:
\begin{equation}
\begin{aligned}
\delta \rho_{S+} = - \frac{\delta m_{*8}}{\calC_S^2}, 
\label{Variation_mass_8}
\end{aligned}
\end{equation}
where the variation in $\rho_{S+}$ is obtained using:
\begin{equation}
\begin{aligned}
\delta \rho_{Sp,n} =& 3\left(\frac{\rho_{Sp,n}}{m_{*p,n}}- \frac{\rho_{p,n}}{\mu_{*p,n}} \right)  \delta m_{*p,n} \\
&+\quad \frac{m_{*p,n}}{\mu_{*p,n}} \frac{k_{Fp,n}^2}{\pi^2} \, \delta k_{Fp,n}.
\end{aligned}
\end{equation}
After some algebra and substituting the variations in $\mu_{*3}$ and $\mu_{*8}$ from \eqref{Chemical_potential_variations_3} and \eqref{Chemical_potential_variations_8}, we find:	
\begin{equation}
\begin{aligned}
3&\left(\frac{\rho_{S+}}{m_{*8}}- \frac{\rho_{+}}{\mu_{*8}} + \frac{1}{\calC_S^2} \right)\,\delta m_{*8} \\
&\quad + \frac{m_{*8}}{\mu_{*8}} \delta \rho_{+} - \frac{m_{*8} \pi^2}{\mu_{*8}^3\,k_F } \frac{\delta \rho_{-}^2}{4} = 0.
\label{Mass_variation}
\end{aligned}
\end{equation}
We must also enforce the vanishing of the Fermi pressure.  The first order variation of the Fermi pressure vanishes identically.  The second order term is:
\begin{equation}
\begin{aligned}
\delta P^N_2 =&\frac{1}{4} \left (\rho_{+} \delta \mu_{*8} + \mu_{*8} \delta \rho_{+} \, + \, \delta \mu_{*3}  \delta \rho_{-}\right) + C_V^2 \rho_{+} \delta \rho_{+} \\
&- \frac{\overline{\calC_{V}^2}}{4} \delta \rho_{-}^2
+\frac{1}{4 C_S^2} \left(3 m_N -2 m_{*8}\right) \delta m_{*8}.
\label{Pressure_variation}
\end{aligned}
\end{equation}
After using \eqref{Chemical_potential_variations_3} and \eqref{Chemical_potential_variations_8}, the zero pressure equation becomes:
\begin{equation}
\begin{aligned}
&\left(\frac{3 m_N -2 m_{*8}}{4 C_S^2} + \frac{\rho_{+} m_{*8}}{2\, \mu_{*8}} \right) \delta m_{*8} \\
&+\biggl( \frac{\mu_{*8}}{4} + C_V^2 \rho_{+} +\frac{k_F^2}{12 \, \mu_{*8}} \biggr) \delta \rho_{+} \\ 
&+\biggl( \frac{\pi^2 \, (5 m_{*8}^2 - 4 k_F^2)}{48 \, k_F\, \mu_{*8}^3}
-\frac{\overline{\calC_{V}^2}}{4} \biggr) \delta \rho_{-}^2 = 0.
\label{Pressure_variation_2}
\end{aligned}
\end{equation}
Equations \eqref{Mass_variation} and \eqref{Pressure_variation_2} are solved to express $\delta m_{*8}$ and $\delta \rho_{+}$ in terms of $\delta \rho_{-}^2$.
To determine $\overline{\calC_{V}^2}$ we need the second variation in the energy density $\mathscr{E}$.  This quantity is discussed below.

\subsection{Calibration of $\overline{\calC^2_{V}}$}
\label{CV200BAR}
\begin{figure}[htp]
\includegraphics[trim={0 0 0 0},scale=.8]{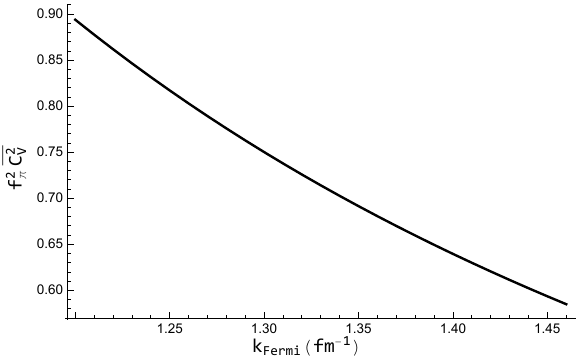}
\centering
\caption{Plot of $\overline{C_{200}^V} \equiv f^2_\pi$ $\overline{\calC}_{V}^2$ against the Fermi level in fm$^{-1}$.  The behavior is roughly linear in the range considered and corresponds to a one-third power of the number density.}
\centering
\label{Mean_CV200_Versus_kFermi}
\end{figure}

We start with the vanishing of the pressure and the relationship:
\begin{equation}
\begin{aligned}
\epsilon^N+P^N =& \mu_{p} \rho_{p}  +  \mu_{n} \rho_{n} = \mu_8 \, \rho_+ +  \mu_3 \, \rho_-,
\end{aligned}
\label{Energy_pressure_equation}
\end{equation}
where
\begin{equation}
\begin{aligned}
\mu_8 = &  \mu_{*8} + C_V^2 \rho_{+} \, , \\
\mu_3 = &  \mu_{*3} - \tfrac{1}{2} {\overline{\calC^2_{V}}} \rho_{-} \,. 
\end{aligned}
\label{Chemical_potential_equations}
\end{equation}
The zeroth order energy density when $Z=N$ follows at once:
\begin{equation}
\begin{aligned}
\epsilon^N_0 = &  \mu_{*8} \, \rho_+ + C_V^2 \,\rho_+^2\,.
\end{aligned}
\label{Zeroth_order_energy_equation}
\end{equation}
The first order energy term vanishes.  The second order term is:
\begin{equation}
\begin{aligned}
\delta \epsilon^N_2 = & \rho_+ \, \delta \mu_{*8} + \mu_{*8} \, \delta \rho_+ + \delta \mu_{*3} \, \delta \rho_{-} \\
&+ 2 C_V^2 \,\rho_+ \, \delta \rho_{+} - \tfrac{1}{2} \overline{\calC_{V}^2} \, \delta \rho_{-}^2 \, .
\end{aligned}
\label{Second_order_energy_equation}
\end{equation}
Finally, we can express $\delta \rho_{-}$ in terms of the relative neutron excess as:
\begin{equation}
\begin{aligned}
\delta \rho_- = \frac{Z-N}{Z+N}\, \rho_+.
\end{aligned}
\end{equation}
\noindent
The parameter $\overline{\calC_{V}^2}$ can be calibrated in two ways.  In the first, we merely ascribe all of the second order energy to the asymmetry term in the liquid drop formula \eqref{Semi_Empirical_Mass_Formula}
 for $\overline{\calC_{V}^2}$:
\begin{equation}
\begin{aligned}
\delta \epsilon^N_2 = a_{Asym} \left(\frac{Z-N}{Z+N}\right)^2  \, \rho_+  = a_{Asym} \frac{\delta \rho_-^2}{\rho_+}
\end{aligned}
\end{equation}
where $a_{Asym}$ is fit to SEMF observation.  In the second approach, we calibrate directly to the binding energies of isotopes, possibly using the liquid drop formula to correct for effects that we have ignored in this paper such as the Coulomb and surface terms.  Both approaches give comparable results.  Figure \ref{Mean_CV200_Versus_kFermi} shows the behaviour of $\overline{\calC_{V}^2}$ for different values of $k_F$.
\end{document}